\documentclass[10pt, twocolumn, journal,web]{article}
\usepackage{times}
\usepackage[margin=16mm]{geometry}
\usepackage{amsmath,amssymb,amsfonts}
\usepackage{subfig}
\usepackage[utf8]{inputenc}
\usepackage{utfsym}
\usepackage{multirow}
\usepackage[normalem]{ulem}
\usepackage{amsmath,amssymb,amsfonts}
\usepackage{algorithmic}
\useunder{\uline}{\ul}{}
\usepackage{url}
\def\BibTeX{{\rm B\kern-.05em{\sc i\kern-.025em b}\kern-.08em
    T\kern-.1667em\lower.7ex\hbox{E}\kern-.125emX}}
\usepackage{amssymb}
\usepackage{cite}
\usepackage{graphicx}
\usepackage{authblk}

\usepackage{fancyhdr}
\fancypagestyle{plain}{%
  \fancyhf{}%
  \fancyfoot[R]{\thepage}%
  \lhead{This work has been submitted to the IEEE for possible publication. Copyright may be transferred without notice, after which this version may no longer be accessible.}
}

\makeatletter
\renewcommand\AB@affilsepx{, \protect\Affilfont}
\makeatother
\usepackage{tabularx}
\providecommand{\keywords}[1]
{
  \small	
  \textbf{\textit{Keywords---}} #1
}

\begin{document}

\title{\textbf{Vision Transformer-based Model for Severity Quantification of Lung Pneumonia Using Chest X-ray Images}}
\author[1]{Bouthaina Slika}
\author[1, 2, 3, *]{Fadi Dornaika}
\author[4, 5]{Hamid Merdji}
\author[4, 6, *]{Karim Hammoudi} 
\affil[1]{\textit{University of the Basque Country}}
\affil[2]{\textit{IKERBASQUE}}
\affil[3]{\textit{Ho Chi Minh City Open University}}
\affil[4]{\textit{University of Strasbourg}}
\affil[5]{\textit{INSERM, Hôpital Univ. de Strasbourg}}
\affil[6]{\textit{Université de Haute-Alsace, IRIMAS}}
\affil[*]{\textit{IEEE Member}}
\affil[ ]{

\small\texttt{bslika001@ehu.eus, fadi.dornaika@ehu.eus, merdji.hamid@gmail.com, karim.hammoudi@uha.fr}}
\date{}
\maketitle
\begin{abstract}
To develop generic and reliable approaches for diagnosing and assessing the severity of COVID-19 from chest X-rays (CXR), a large number of well-maintained COVID-19 datasets are needed. Existing severity quantification architectures require expensive training calculations to achieve the best results. For healthcare professionals to quickly and automatically identify COVID-19 patients and predict associated severity indicators, computer utilities are needed. In this work, we propose a Vision Transformer (ViT)-based neural network model that relies on a small number of trainable parameters to quantify the severity of COVID-19 and other lung diseases. We present a feasible approach to quantify the severity of CXR, called Vision Transformer Regressor Infection Prediction (ViTReg-IP), derived from a ViT and a regression head. We investigate the generalization potential of our model using a variety of additional test chest radiograph datasets from different open sources. In this context, we performed a comparative study with several competing deep learning analysis methods. The experimental results show that our model can provide peak performance in quantifying severity with high generalizability at a relatively low computational cost. The source codes used in our work are publicly available at  \url{https://github.com/bouthainas/ViTReg-IP}.
\end{abstract}

 \hspace{10pt}
\section{Introduction}
\label{Introduction}

The number of deaths caused by coronavirus disease-19 (COVID-19) continues to rise even after vaccination by mandatory policies in most countries \cite{world,ng2020imaging}. Many physicians have turned to new tactics and technologies due to the increased impact of the pandemic on healthcare systems around the world. Chest radiographs X-rays (CXR) offer a relatively noninvasive method to track disease progression \cite{Chest,ng2020imaging}. CXR imaging is becoming more popular and more widely used worldwide, as demonstrated by many recent studies \cite{mao2020assessing,guan2020clinical, toussie2020clinical, jacobi2020portable, rubin2020role, huang2020clinical, wong2020frequency}.
CXR imaging devices are more widely accessible than CT scanners due to their lower cost and faster decontamination times \cite{gietema2020ct}. In addition, because portable CXR units are available, imaging can be performed within a stationary unit, which significantly reduces the risk of contamination transmission\cite{rubin2020role,jacobi2020portable,dennie2020canadian}. Finally, CXR imaging in patients with respiratory complaints is considered a commonly accepted good standard practice in medicine \cite{nair2020british} and it has been shown to provide insightful information about disease progression \cite{wong2020frequency}.
Numerous studies have looked at CXR images, particularly those of SARS-CoV-2-positive patients \cite{guan2020clinical,huang2020clinical,kong2020chest}, with bilateral abnormalities, ground-glass opacity, and interstitial abnormalities. Determining the severity of a patient's disease is an important help of CXR assessment by physicians to guide the treatment and management of the disease, as it relies on the detected imaging features and observation of their progression and extent over the duration of disease onset. As a result, several recent studies have focused on severity scoring to quantify the severity of lung disease \cite{wong2020frequency, toussie2020clinical}. Disease severity can help physician to determine the appropriate treatment and monitoring for each patient. Radiology services often employ experienced physicians for whom determining the severity of a CXR is not an easy task. Clinical diagnosis with the help of a computer could make this difficult task easier for physicians. In this research, we developed and studied a model that can predict the severity of lung pneumonia based on CXR and can be used to support patient care management. Escalation or de-escalation of care, particularly in the intensive care unit (ICU), may be based on the ability to assess the severity of pulmonary infection. Over time, a patient’s response to treatment and disease progression can be objectively and quantitatively tracked using an automated method. We anticipate that the use of CXRs from a global pool of patients with pulmonary infections and normal patients can lead to a reliable and generic computer-aided severity grading of lungs. Throughout the study, we are interested in investigating the performance of our proposed model in predicting a scalar representing severity, rather than just classifying images as infected or uninfected. Recent work has shown that Deep Learning can be used to solve regression problems such as estimating the age of faces \cite{Dornaika2020}, predicting the beauty of faces \cite{BOUGOURZI2022}, and evaluating the risk score of breast cancer disease progression \cite{breastcancerrisk}.

Then our task is a regression task where we need ground truth scores for supervised learning. Specifically, in this study, we develop, train, and validate a transformer-based deep neural network capable of achieving the required score prediction. It will be used to perform multiple scoring systems by exploiting CXRs from both infected and normal patients. In this way, we can assess the feasibility of computer-aided lung severity scoring towards assistance to support accurate diagnosis and treatment. Although transformer-based architectures have been widely used recently \cite{Li2023 }, most research focuses on solving a classification problem rather than a regression problem as in the case of our research.

Our main contributions are summarized below:
\begin{itemize}
\item Formulation of a generalized and outperforming approach based on a vision transformer (ViT) to predict the severity of a lung infected with COVID-19.
\item Derivation of mixing and fusing data augmentation methods, originally developed for classification tasks, as a scoring augmentation stage for our regression solution to generate a larger dataset.
\item Carrying out a comparative study by  exploiting \textit{state-of-the-art} databases (RALO, Brixia, Danilov \textit{et al.} COVID-19 and Cohen COVID-19) and eight different deep learning models (COVID-NET, COVID-NET-S, ResNet50, InceptionNet, XceptionNet, Swin Transformer, MobileNetV3, and Stonybrook Feature Extraction).
\item Conducting a series of  ablation studies showing the relative contribution of each component in  our ViTReg-IP.
\item We made our source codes publicly available to encourage other researchers to use them as a benchmark for their studies: \url{https://github.com/bouthainas/ViTReg-IP}.
\end{itemize}

The remainder of the paper is organized as follows: Related studies and a review of the \textit{state-of-the-art} were described in Section \ref{Relative work}. The description of our proposed generalized pneumonia severity quantification model is presented in Section \ref{Proposed methodology}. The performance evaluation, including the datasets used and the experimental results, as well as a detailed evaluation of the performance of each approach in severity assessment, are presented in Section \ref{Performance Evaluation}. In Section \ref{Discussion}, we interpret and discuss the obtained results. Section \ref{conclusion} summarizes the results and provides some concluding notes.
\begin{figure*}[ht!]
	\includegraphics[width=5in]{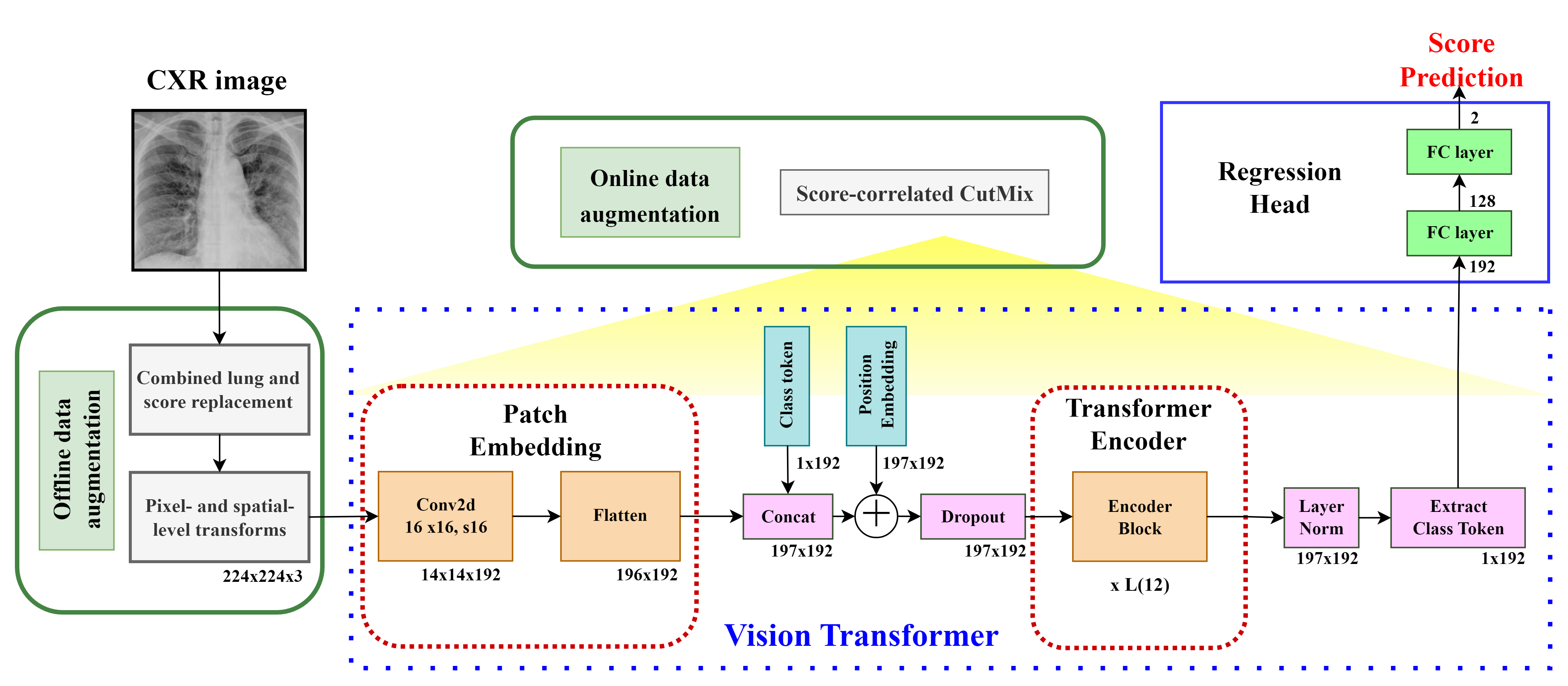}
	\centering
	\caption{ Overview of the proposed ViTReg-IP model.}
	\label{fig: ViTReg-IP}
\end{figure*}
\section{Related Work}
\label{Relative work}
COVID-19 has monopolized the focus and financial resources of researchers in digital technologies, artificial intelligence, and data science from the onset of the pandemic \cite{ting2020digital,latif2020leveraging}. Shi \textit{et al.} \cite{shi2020review} and Islam \textit{et al.} \cite{islam2021state} state that there are several artificial intelligence techniques in image data acquisition, segmentation, and diagnosis for COVID. The authors organized prior work according to different tasks, such as contactless imaging workflow, image segmentation, disease detection, radiological feature extraction, and severity quantification. At the very beginning of the outbreak, when a systematic collection of a large CXR dataset for deep neural network training was still difficult, Oh \textit{et al.} \cite{oh2020deep} proposed the use of convolutional neural networks (CNN) to analyze CXRs for presumptive early diagnosis and better patient handling based on signs of pneumonia. Researching deep learning techniques was also investigated in \cite{hammoudi2021deep} for autonomous assessment of CXR images to provide healthcare with accurate tools for COVID-19 screening and patient diagnosis. The increasing availability of CXR datasets from patients with COVID-19 during the outbreak has focused almost all research efforts on diagnosis-oriented image interpretation investigations. There are so many studies that apply AI techniques to the acquisition, segmentation, and classification of imaging data for COVID-19, whether using CXRs or CT scans, that it would be difficult to cover them all \cite{shi2020review,covidPred2021}. We refer only to those related to our work and the most recent emergent challenges. 

Even though CXR imaging modality is frequently used in many healthcare facilities, few studies to date have presented AI-driven solutions for COVID-19 disease surveillance and pneumonia severity assessment based on CXR especially those that predict a score, as is done in our study. 
The first paper was by Irmak \textit{et al.} \cite{irmak2021covid} was based on a quantitative CXR assessment \cite{orsi2020feasibility}. However, this study needs readers with more knowledge to confirm the consistency of the severity score. After that, the  COVID-Gram \cite{liang2020development} and the deep learning applied works of Liang \textit{et al.} \cite{liang2020early} were used for COVID-19 identification based on CXR abnormality. The extent of lung pneumonia was determined in the work of Colombi \textit{et al.} \cite{colombi2020well} to indicate the severity of the disease. Another notable work was COVID-NET-S \cite{wong2020covid}, one of the earliest research projects on COVID-19 severity assessment, in which the author developed a deep neural network to forecast extent scores from CXR. To do this, they had to train their model on a huge dataset. Several features from a neural network that has been trained on CXR datasets other than COVID-19 are taken into account for their predictive score on the estimate of COVID-19 severity ratings in \cite{cohen2020predicting}. Ridley created a unique type of deep learning algorithm called Convolutional Siamese Neural Network (CSNN) to produce a score of COVID-19 patient pulmonary X-ray severity (PXS) that was well correlated with radiologist assessments and could also be used to predict whether or not a patient will need to be incubated before death \cite{ridley2020ai}. In \cite{li2020automated}, a transfer learning method is applied from a large non-COVID-19 dataset to a small COVID-19 one to show a clear relationship between a lung-based severity score rating and automated prediction. Improved generalizability was attained in a subsequent study by the same authors \cite{li2022multi}. In \cite{frid2021covid}, pneumonia localization and lung segmentation networks were used to produce a geographic extent severity score that was annotated and linked with experts' evaluations on 94 CXRs. The geographic extent and opacity severity scores were predicted in \cite{wong2021towards}. The authors in \cite{wong2020covid} used a modified COVID-19 detection architecture  and stratified Monte Carlo cross-validation. These were performed on 396 CXRs, measuring the relationship with respect to expert annotations. CheXNet, which was pretrained on ImageNet and later trained to predict COVID-19 severity using a unique dataset, was proposed in the study by Kwon \textit{et al.} \cite{kwon2020combining}. An end-to-end deep learning architecture was used in \cite{signoroni2021bs} to predict  a multi-regional score; Brixia score based on  CXR images (CXR), indicating the severity of lung damage in COVID-19 patients. The architecture used a large dataset and needed several pieces of training for segmentation and subsequent prediction.

When deep architectures are used to evaluate datasets with thousands of images, the computational cost is often very high. Furthermore, when datasets are used to represent only COVID-19 images and neglect other types of pneumonia, we may encounter problems that lead to inefficient results. These early studies confirm the need for specialized methods for this difficult visual task, the urgency of working with small annotated datasets, and the need for computationally inexpensive models for meaningful explanatory solutions while demonstrating the feasibility of COVID-19 severity estimation on CXRs. In contrast to the aforementioned methods, our work shows how a focused straightforward technical solution that can handle the complexity of an organized severity assessment can achieve high performance and robustness at a low computational cost.
\section{Proposed Methodology}
\label{Proposed methodology}
\subsection{Combined Feature Extraction Models}
Given a CXR image with  dimensions $H\times W\times C$, where $H\times W$ is the spatial resolution and $C$ is the number of channels, our goal is to predict the respective lung infection severity score. In our approach, a regressor is backed by the vision transformer ViT \cite{dosovitskiy2020image}. Figure \ref{fig: ViTReg-IP} shows the schematic diagram of our proposed model ViTReg-IP.
\subsubsection{Vision Transformer Backbone}
In this study, our ViTReg IP model is based on the vision transformer \cite{dosovitskiy2020image}. The parameters are initialized using the deep neural network, which has already been pre-trained on ImageNet \cite{voulodimos2018deep}. The non-hierarchical ViT design of the deep neural network was used in this study to underpin the architecture of the proposed model used to evaluate the functionality of the computational severity of lung disease. The ViT reshapes the input CXR image into a sequence of flattened 2D patches, where each patch has size \(P\)$\times$\(P\) and \(N = \frac{H\times W}{P^2}\) is the number of image patches. Using a trainable linear projection, we translate the vectorized image patches $\textbf{x}_p \in \mathbb{R}  ^{P^2 \times C }$ into a latent $D$-dimensional embedding space. We learn certain position embeddings that are added to the patch embeddings to obtain the position information that encodes the spatial patch information.  Thus, the encoding of the $N$ patches  is represented   by  the $N \times  D$ matrix $\textbf{z}_0$ as follows:
\begin{equation}
\textbf{z}_0 = [\textbf{x}^1_p \textbf{E}; \textbf{x}^2_p \textbf{E};...;  \textbf{x}^N_p \textbf{E}] + \textbf{E}_{pos},\label{eq1}
\end{equation}
where $\textbf{E}  \in \mathbb{R}  ^{P^2 \times C \times D }$ is the patch embedding projection, and $\textbf{E}_{pos}$ is the position embedding.
The Multihead Self-Attention (MSA) and Multi-Layer Perceptron (MLP) blocks are included in all  $L$ layers of the transformer encoder \eqref{eq2} and \eqref{eq3}. Consequently, the following can be expressed as the output of the $l^{th}$ layer:
\begin{equation}
{\textbf{z}'_{l} = MSA (LN(\textbf{z}_{l-1}))+\textbf{z}_{l-1}}, \label{eq2}
\end{equation}
 \begin{equation}
\textbf{z}_{l} = MLP(LN(\textbf{z}'_{l}))+\textbf{z}'_{l}, \label{eq3}
\end{equation}where $LN ( . ) $ represents the layer normalization operator and $\textbf{z}_{l}$ is the encoded image representation at layer $l$. The structure of a transformer encoder is illustrated in Figure \ref{vit encoder}. In our tests, we use a tiny ViT backbone with $(W, H) = (224,224)$, $C = 3$, $P = 16$, $L = 12$, and $D = 192$.
\subsubsection{Regression Head}
\begin{figure}[h!]
\centering\subfloat[Representation of the vision transformer blocks.]{\includegraphics[width=\columnwidth]{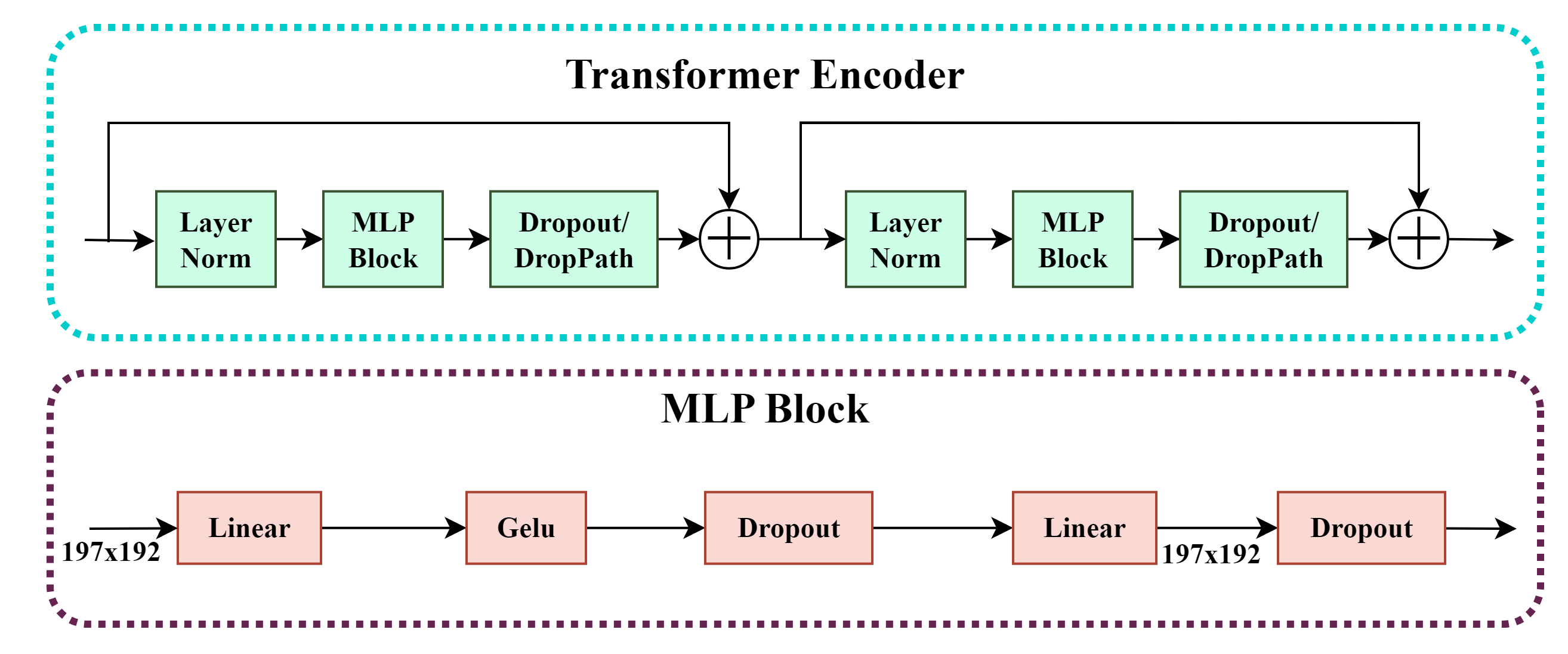}\label{vit encoder}}\\
\subfloat[Representation of the regression head.]{\includegraphics[width=2.5in]{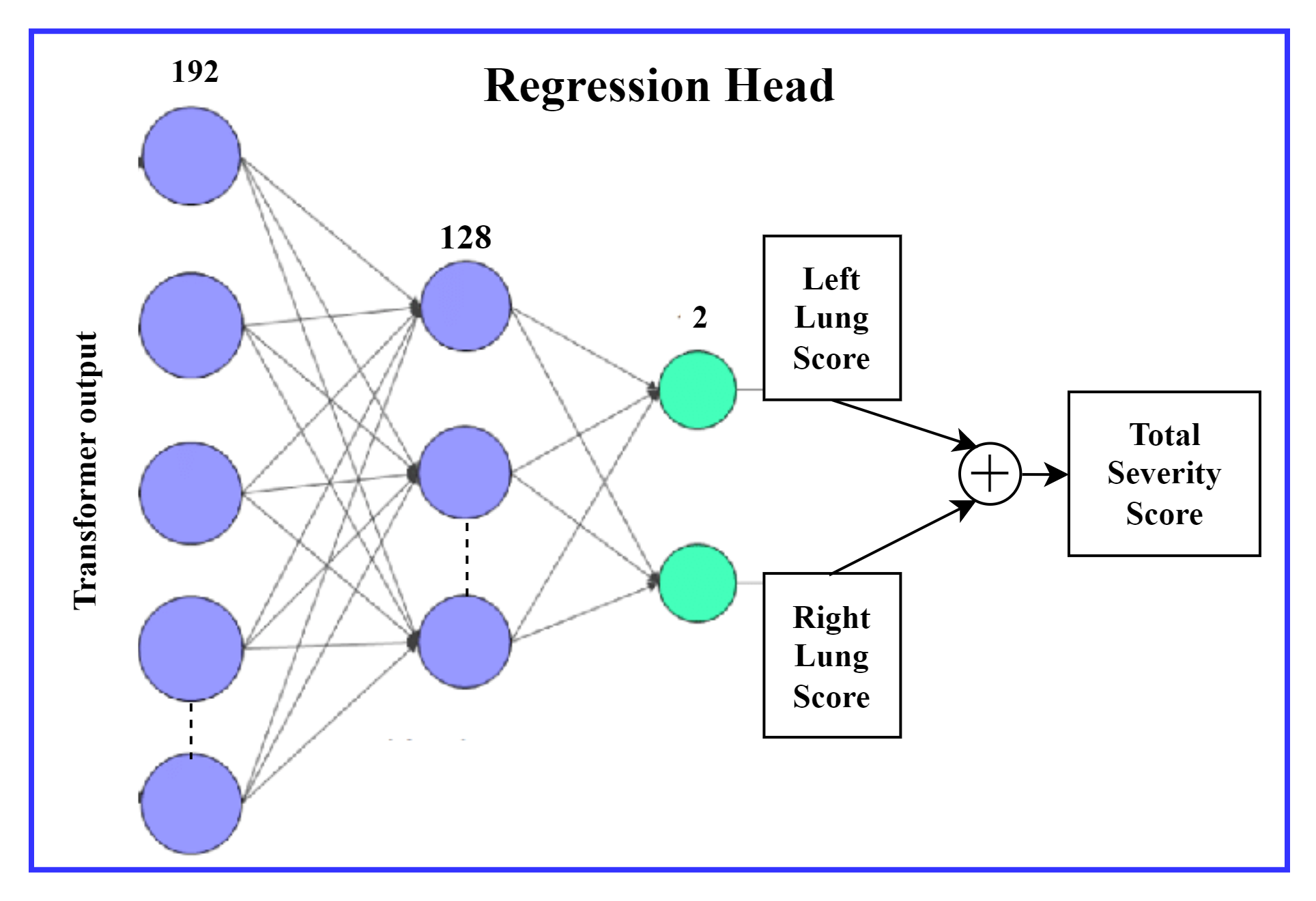}\label{reg head}}
\centering
\caption{Detailed representation of the feature extraction blocks and the regression block in the proposed method.}
    \label{fig: blocks}
    \end{figure}

Our regressor consists of two Fully Connected (FC) layers. It takes as input the CLS token  provided by the last layer of the  ViT backbone. It then computes two predictions (left and right lungs) for the score since the infection levels in the two lungs are independent of one another. The regression head is thus composed of two new fully connected layers that replace the  ViT's classification layer. These two layers are: a linear layer with 128 neurons and a linear layer with 2 neurons. An illustration of the regression head is shown in Figure \ref{reg head}. To determine the extent of infection, each score serves as a prediction for the severity of the left lung and the right lung. The final output of the network is the score prediction with a value falling between 0 and 8, which is the sum of the two output scores.
\begin{equation}
\textbf{p} = [p_l; p_r ]  = FC_2(FC_1(\textbf{CLS}_\textbf{L})), \label{eq4}
\end{equation}
where $FC_1$ and $FC_2$ are the two trainable fully connected layers respectively, $\textbf{CLS}_\textbf{L}$ is the $\textbf{CLS}$ token of the final transformer layer, and $\textbf{p}$ is the predicted vector including the left and right lung scores. The final global output score is the sum of these two scores, $p= p_l +p_r$.
The final prediction of our VitReg-IP model, $p$, corresponds to the predicted CXR severity. Using a specific loss function, this value is then compared to the ground truth score to train the entire network. The radiological scores are the ground-truth scores. It is important to note that although our proposed model produces predicted scores for each lung, the training data don't necessarily need to include ground-truth scores for each lung because the loss function is based on the total score.
\subsection{Data Augmentation}
All CXR data used in this study underwent data normalization, a crucial step that ensures that each input parameter pixel has a uniform data distribution. This accelerates convergence during training of the model. In addition, to facilitate the training of the deep neural networks in this study, all CXR data were reformatted to identical data dimensions of size $224 \times 224 \times3$.
We construct our deep neural network by applying successive operations in order to convert the CXR input data into the projected severity scores (e.g., geographic extent score, opacity extent score, and Brixia score). The efficiency and effectiveness of our network are highly dependent on the accessibility to data as well as the preparation of training and test data.

If learned weights perform well in the training set but poorly in the test set, these models are overfitted. In the context of this study, we need to expand the size of the dataset used to avoid overfitting, which prevents generalization of the model. Indeed, in precision health, there is often a lack of input data due to the novelty of the tackled topics and the high cost of labeling by medical experts \cite{willemink2020preparing}. In our architecture, the size of CXR images is increased by operating dropping and merging data augmentation methods.

More specifically, the data augmentation in this study involves the creation of new training images  from the original  CXR data of the training set using the combined offline lung and score replacement (inspired by the lung replacement method \cite{ridzuan2022challenges}) and the online score-correlated CutMix derived from the simple CutMix \cite{yun2019cutmix}. These two augmentation methods increase the data variety and potential of deep neural networks in terms of robustness and accuracy. The above two augmentation methods were developed and adapted for our regression problem and thus are used to generate the augmented images as well as their corresponding ground-truth scores.
\subsubsection{Combined Lung and Score Replacement}
\begin{figure}[h!]
\centering
\includegraphics[width=\columnwidth]{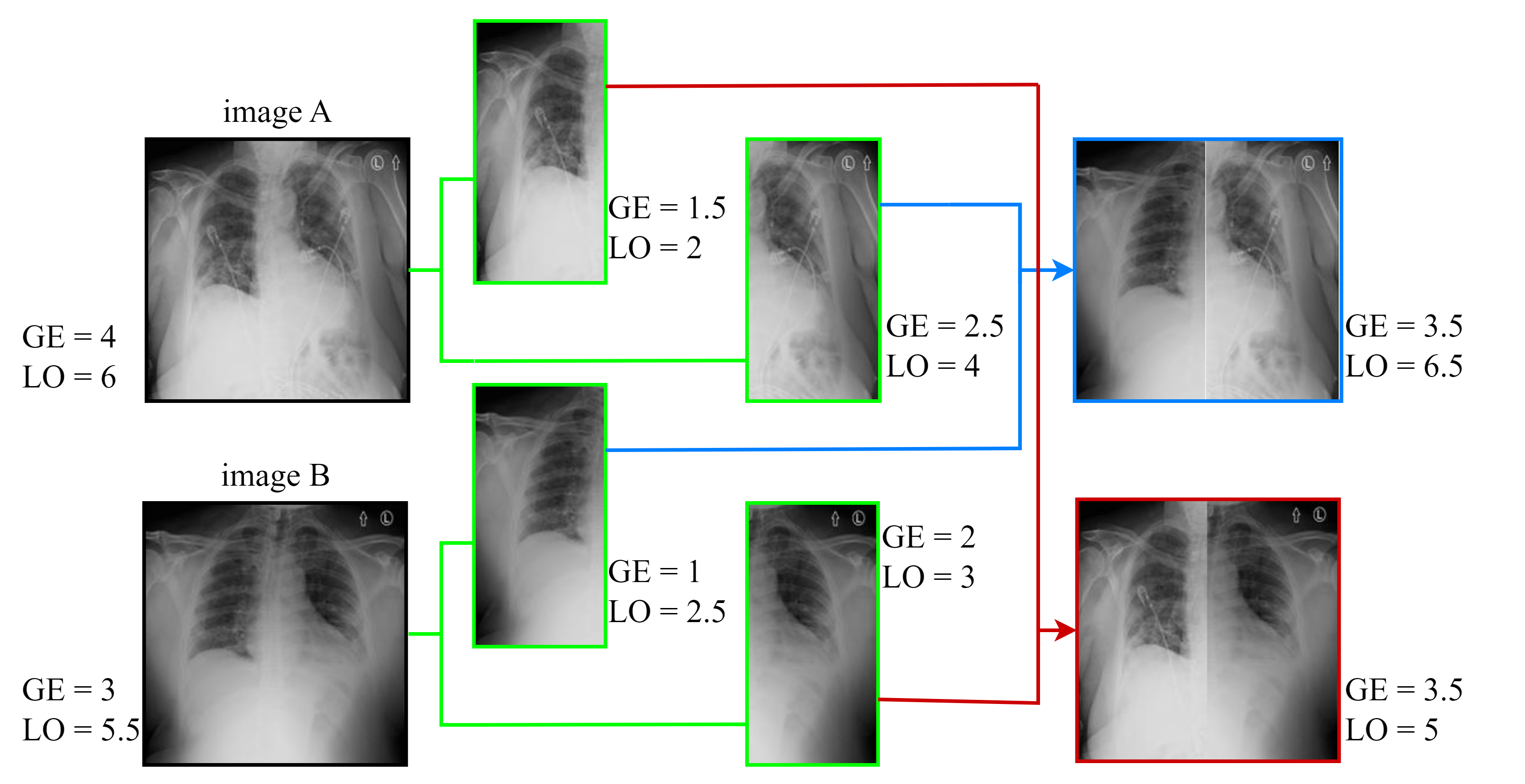}\\
\caption{Combined lung and score replacement method applied on CXR images.}
\label{fig: lung rep}
\end{figure}

We applied a lung replacement procedure already proposed for a  classification problem, but in an improved version for the case of regression. The principle is based on replacing the left or right lung of a given patient with the opposite lung from a CXR of another patient. Lung replacement was applied to CXRs of the same class in \cite{ridzuan2022challenges} to increase the training data. However, since we have a regression problem  in our case, we can apply the lung segmentation to any two CXR images. Moreover, the new ground truth scores of the two resulting images are calculated by adding the individual lung scores of the left and right parts of the original images. Thus, we apply a combined lung and score replacement. Thus, the individual lung images are merged and the ground truth scores of the combined lungs are added to also determine the ground truth scores of the created images. To use this method, the individual ground truth scores must be available. An illustration of the process can be seen in Figure \ref{fig: lung rep}. We applied this method to the training portion of the dataset, which originally consisted of 1878 images for the RALO dataset \cite{wong2021towards}. A combined lung and score replacement was applied to these images, resulting in two sets of synthetic images since we have two types of severity scores. The new training dataset now consists of a total of 5634 CXR images.
\subsubsection{Score-correlated CutMix}
\begin{figure}[h!]
 \includegraphics[width=2.5in]{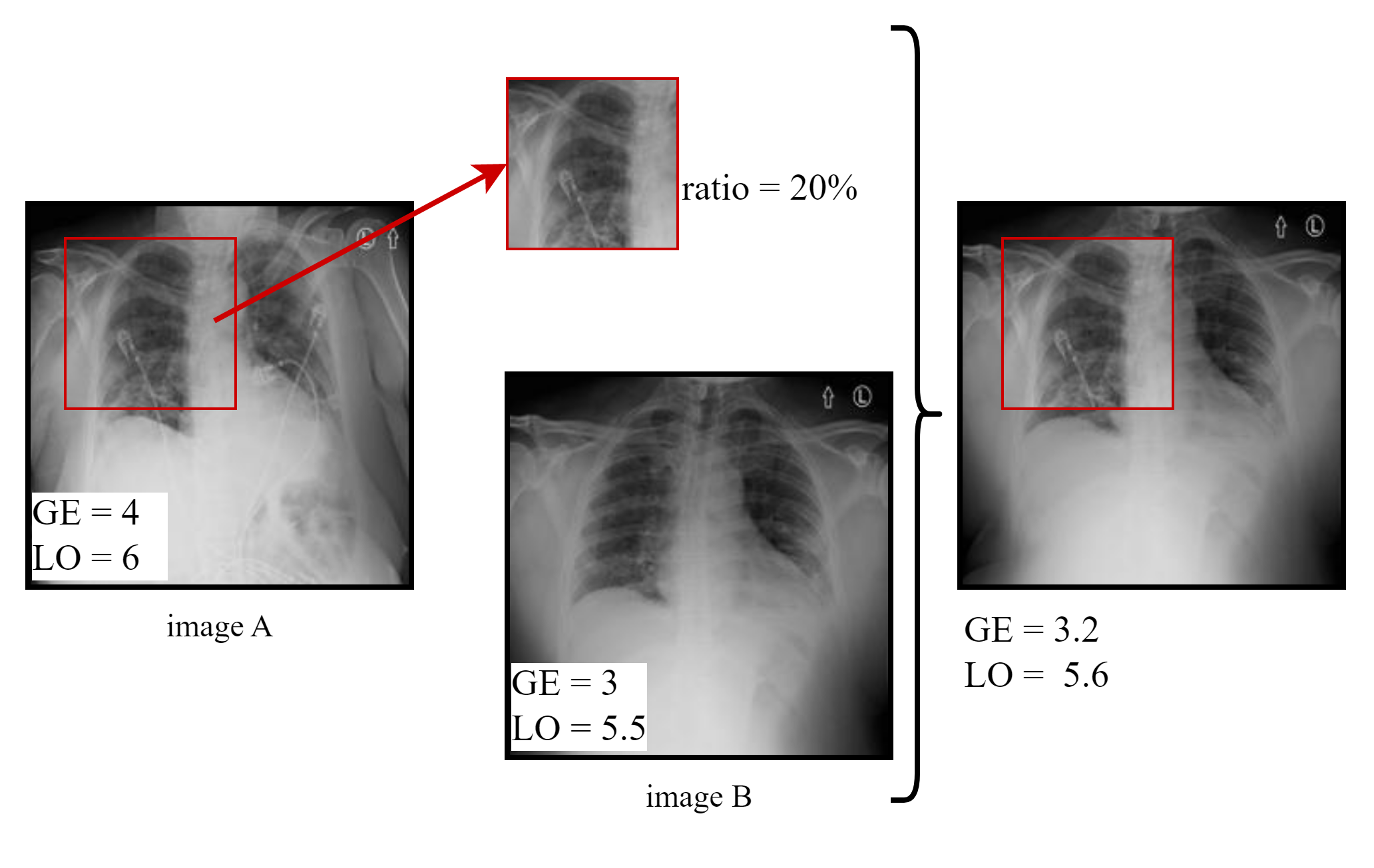}
\centering
\caption{Score-correlated CutMix method applied on a CXR image.}
\label{fig: cutmix}
\end{figure}

The CutMix method creates a locally natural image by replacing a single image region with a patch from another training image. We use CutMix by replacing cropped portions from image $A$ with a patch from another image $B$ in which the size is chosen randomly within a given range, to create a new image, as shown in Figure \ref{fig: cutmix}. This method was used during online training, where one image is CutMix-ed with other images from the same batch in each epoch.
In our study, we have a regression problem, i.e., we are interested in computing the new ground truth score of the synthesized image. Therefore, we adapt the conventional CutMix (purely image-based) to our regression problem and create a score-correlated CutMix. According to the total number of pixels in the fused images, the ground truth labels are computed by a weighted average   of the ground truth scores of both images, as shown in \eqref{eq5}.
\begin{equation}
\overline{y} = \lambda*y_A + (1-\lambda)*y_B,\label{eq5}
\end{equation}
where \(\lambda\) is the ratio between the size of the replaced area and the total size of the image, \(y_A\) and \(y_B\) are the ground truth scores of images $A$ and $B$, respectively, and \( \overline{y}\) is the new ground truth score.

Efficient deep network training requires a significant amount of data. From a few images, it is possible to generate many more augmented images using the combined lung and score replacement method. Additionally, using the score-correlated CutMix during model training enables the creation of diverse images and their corresponding relative scores.
\subsection{Loss Function and Optimizer}
$L_1$ loss, also known as Absolute Error Loss, is the loss function we use for training.  The used loss function is described in \eqref{eq6}. 
\begin{equation}
{\cal {L}}=  \sum_{i=1}^{N_b}   |p_i-\hat {p}_i|,
\label{eq6}
\end{equation}
where \({p}_i\)  and  \(\hat {p}_i\) are   the ground truth score and  the predicted score of the $i^{th}$  image, respectively. $N_b$ is the size of the batch.
The experiments' findings demonstrated that when trained on the analyzed RALO dataset of CXR data using $L1$ Loss, ViTReg-IP had the best performance in comparison to other loss functions. 

The stochastic gradient descent (SGD) optimizer is employed to change the model's parameters. Its role is to modify model weights in order to minimize the loss function. This optimizer is chosen after testing varied optimizers and comparing the model's performance.
\section{Performance Evaluation}
\label{Performance Evaluation}
\subsection{Datasets }
The main objective of this study is to evaluate the feasibility of Deep Learning-based computer assistance in assessing the severity of lung disease. To this end, we evaluate our solution as well as other deep neural networks  capable of assessing CXR images of patients with different severities of lung infections. To this end, several CXR datasets were used in this study
\cite{wong2020covid, Savardi_Signoroni_Benini_Adami_Farina_2021, coviddata2022, cohen2020covid}. Table \ref{tab:data summary} summarizes the datasets used.
\begin{center}
	\begin{table}[h!]
		\caption{Summary of severity labeled CXR datasets used in our study.}
      	\centering
		\resizebox{\columnwidth}{0.4in}{
        \begin{tabular}{lrlll}
            \hline
            \noalign{\vskip 0.5mm}  
              \textbf{Dataset}& \textbf{Data Size}&\textbf{Annotations}&\textbf{Score Range}\\
            \hline   
            \noalign{\vskip 0.5mm}  
          	 \textit{RALO}\cite{wong2020covid} &2373&GE, LO&[0-8]\\
            \textit{Brixia}\cite{Savardi_Signoroni_Benini_Adami_Farina_2021} & 4695 &Brixia &[0-18]  \\
            \textit{Danilov et al. COVID-19}\cite{coviddata2022} &1364&COVID score&[0-6]\\
            \textit{Cohen COVID-19}\cite{cohen2020covid}&192+94 &Brixia score+(GE, LO) &[0-18]+[0-8]\\[0.5ex]
			\hline
		\end{tabular}}
	\label{tab:data summary}
	\end{table}
\end{center}

\subsubsection{RALO dataset}

We used the Radiographic Assessment of Lung Opacity Score (RALO) dataset in our study \cite{wong2020covid}. Stony Brook Medicine recorded and scored the RALO dataset to provide researchers with a common COVID-19 dataset. Two renowned radiologists assessed the dataset, which consists of 2373 CXRs, to perform an additional COVID-19 severity analysis. The RALO dataset is divided into training and test data with 1878 and 495 images, respectively. Geographic extent (GE) and lung opacity (LO) are the two assessment metrics considered in the radiological evaluation. The right and left lungs are assessed independently, and the geographic extent of lung involvement by morning opacification or consolidation is scored as follows: 0 = no involvement; $1 = 25\%; 2 = 25-50\%; 3 = 50-75\%$; and $4 =\> 75\%$ involvement. The total score for geographic extent (right + left lung) ranges from 0 to 8 after the scores were summed. The degree of opacity was scored for the right and left lungs separately as follows: 0 = no opacity; 1 = ground glass opacity; 2 = mixture of consolidation and ground glass opacity (less than 50\% consolidation); 3 = mixture of consolidation and ground glass opacity (more than 50\% consolidation); and 4 = complete opacification. The total score for the extent of opacity, obtained by adding the scores for the right and left lungs, ranges from 0 to 8 points \cite{wong2021towards}.
\subsubsection{Brixia dataset}
The Brixia dataset, compiled from a dataset of 4695 CXR images matching the number of images acquired for patient monitoring in ICUs during the pandemic, was one of three datasets used to perform our tests \cite{Savardi_Signoroni_Benini_Adami_Farina_2021}. The following annotations describe the relative Brixia score. The lungs are divided into six zones, three for each lung, when viewed from the anteroposterior (AP) or posteroanterior (PA) angle. Depending on the type and severity of lung abnormalities, a score of 0 (no abnormalities), 1 (interstitial infiltrates), 2 (interstitial and alveolar infiltrates, interstitial dominance), or 3 (interstitial and alveolar infiltrates, alveolar dominance) is assigned for each zone. The six scores can be combined to obtain a Global Score  ranging  from 0 to 18.
\subsubsection{Cohen COVID-19 dataset}
The dataset of Cohen \textit{et al.} COVID-19 is also used \cite{cohen2020covid}. This collection consists of CXR images collected from numerous locations around the world, at different spatial resolutions, and with other unidentified image quality factors such as modality and window-level settings. Two subsets of this dataset were exploited in our analysis:
\begin{itemize}
\item The CXR labeled with the Brixia scores subset is used. The associated Brixia score annotations for the CXR in this collection were prepared by two experienced radiologists, a certified staff member Rs and a trainee Rj, with 22 and 2 years of experience, respectively. The collected dataset consists of 192 CXRs that were fully annotated using the Brixia scoring system.
\item We also used a retrospective cohort of 94 posteroanterior (PA) CXR images from the COVID-19 imaging dataset, which is available to the general public. Physicians for each patient indicated that they were all COVID-19 positive. Ratings of geographic extent and opacity extent ratings are used to label these images.
\end{itemize}
\subsubsection{Danilov \textit{et al.} COVID-19 dataset}
In this dataset, the authors provide a collection of CXRs from COVID-19-positive and negative patients. There are a total of 1,364 images. Of these, 580 images show COVID-19 positive results (43\%), while 784 images show no results at all (57\%). Each image was assigned a score between 0 and 6, with 0 representing no abnormalities and 6 representing a severe case of COVID-19 involving more than 85\% of the lungs. In addition to the COVID-19 data, it also includes CXR images of healthy lungs without pneumonia or other abnormalities \cite{coviddata2022}.
\subsection{Experimental Setup}
\label{sec}
We compared the performance of our deep neural network with other deep learning architectures and evaluated our ViTReg-IP model against additional datasets to investigate the effectiveness of the deep neural network we developed for the computational assessment of lung disease severity. Several deep-learning models were used for comparison. We show that the proposed network model provides higher sensitivity and interpretability than the current COVID-Net \cite{wong2021towards} and COVID-NET-S\cite{wong2020covid}. We also employ ResNet50, a ResNet variant developed by Kaiming He \textit{et al.} \cite{he2016deep}, with 50 layers, where we replace the output layer with a regression head with two outputs. We also tested the Swin transformer \cite{liu2021swin}. This is a hierarchical transformer architecture whose representation is generated by shifted windows. It can serve as the main structural support for a regression task performed for evaluation. Similarly, the depth-separable regular convolutions of the XceptionNet architecture \cite{chollet2017xception} are put to the test. We also tested the InceptionNet architecture, \cite{szegedy2015going} which emphasizes parallel processing and concurrent feature extraction. Moreover, we tested the model proposed by Cohen \textit{et al.} \cite{cohen2020predicting} which was trained using a large dataset as a feature extractor and allows score predictions. In addition, we tested MobileNetV3, a convolutional neural network tailored to cell phone CPUs through a combination of hardware-aware network architecture search (NAS) and the NetAdapt algorithm \cite{mobilenetv3}. The output of this model was also modified to predict the score through a regression head.

In order to evaluate our experiments, we tested our ViTReg-IP model over several datasets. We trained CXR images of size 224\(\times\)224 each, a batch size of 32, a learning rate of \(1\times10^-3\), and 60 iterations with $L1$ Loss as the loss function. Python language and the PyTorch Lightning learning package were used for the entire model development process.

We compute the mean absolute error (MAE) and Pearson correlation coefficient (PC) between predicted scores outputted by the deep neural networks and ground truth scores by expert radiologists for geographic extent, opacity extent, and Brixia scores in the test sub-set of CXR data for each trial in order to measure the performance of the deep neural networks learned in this study. 
A model evaluation metric used with regression models is the mean absolute error as in \eqref{eq7}.
\begin{equation}
MAE=\frac{\sum_{i=1}^n|y_i-x_i|}{n},\label{eq7}
\end{equation}
where \(y_{i}\) is the prediction and \(  x_{i}\) the true value. The average of the absolute  prediction error over all \(n\) test images is the mean absolute error of a model. 

The linear link between the two sets of scores is gauged by the Pearson correlation coefficient as follows: 
\begin{equation}
PC  = \frac{\sum_{i=1}^{n}  (x_i - \overline{x}) * (y_i - \overline{y}) / n}{\sigma_x * \sigma_y}\label{eq8}
\end{equation}
where $n$ denotes the number of test images, $ y_{i}$  is the predicted score (GE or LO) of the $i^{th}$ test CXR image, $x_{i}$  is its ground-truth score, $ \overline{y}$ is the mean of all predicted scores  and $ \overline{x}$ is the mean of all ground-truth scores. $ \sigma_y$  and  $ \sigma_x $   are  the  standard deviations  of the predicted and  ground-truth  scores, receptively. This correlation coefficient, like others, ranges from $-1$ to $+1$, with $0$ denoting no correlation and $+1$ a precise linear one.
\subsection{Experimental Results and Comparison}
We used nine different models to train the processed and expanded RALO dataset for lung severity assessment: COVID-NET \cite{wong2021towards}, COVID-NET-S\cite{wong2020covid}, ResNet50\cite{he2016deep}, InceptionNet \cite{szegedy2015going}, XceptionNet \cite{chollet2017xception}, Swin Transformer \cite{liu2021swin}, Stonybrook Feature Extraction \cite{cohen2020predicting}, MobileNetV3 \cite{mobilenetv3}, and our ViTReg-IP model. The dataset contains images labeled with the geographic extent and lung opacity, with values ranging from 0 to 8 to denote disease severity, which ranges from normal to severe. The dataset used includes the original images and the images resulting from combined lung and score replacement, and the score-correlated CutMix was applied as an online augmentation method. This applies to all training conducted for all models tested. The models COVID-NET, COVID-NET-S and Stonybrook Feature Extractor were trained unchanged, while the remaining models were used as a backbone to replace the ViT in our proposed model. The results are shown in Tables \ref{tab:ge results} and  \ref{tab:lo results}. For each metric in each table, a thorough investigation of the performance of deep learning models in assessing infection severity is provided. In addition, there are two columns indicating the number of parameters used in each model and the time required for training. Table \ref{tab:ge results} presents the results obtained after training the models using the geographic extent as a label. It shows that our proposed model has the best performance. Table \ref{tab:ge results} similarly shows the results for the score of lung opacity.
\begin{table}[h!]
	\caption{Geographic extent score prediction results.}
	\centering
	\resizebox{\columnwidth}{1in}{
    \begin{tabular}{lrrrr}
    \hline
     \noalign{\vskip 2mm}  
            \textbf{Model}     & \multicolumn{1}{c}{\textbf{MAE $\downarrow$}}    & \textbf{PC $\uparrow$} &\textbf{Number of} &\textbf{Training}\\
                   &   &  &\textbf{parameters} &\multicolumn{1}{c}{\textbf{time}}\\[1ex]
        \hline
         \noalign{\vskip 2mm}  
		 \textit{COVID-NET}\cite{wong2021towards} &4.563&0.545&12 M&40 min\\
		 \textit{COVID-NET-S}\cite{wong2020covid} &4.746&0.581&12 M&40 min\\
         \textit{ResNet50}\cite{he2016deep}  & 1.107&0.684& 23 M&1.5 hr\\
		\textit{Swin Transformer}\cite{liu2021swin} & 0.927& 0.819& 29 M& 2 hr \\
		\textit{XceptionNet} \cite{chollet2017xception} & 0.864&0.802& 23 M &1.5 hr \\		
        \textit{InceptionNet} \cite{szegedy2015going} & 0.717 &0.881 & 24 M&1.5 hr\\		
        \textit{Feature Extraction}\cite{cohen2020predicting} & 0.981 & 0.741& 20 M&1 hr \\
        \textit{MobileNetV3}\cite{mobilenetv3} &0.864&0.822&4.2 M& 40 min\\
        \textit{\textbf{ViTReg-IP (ours) } }&\textbf{0.569}&\textbf{0.923}& \textbf{5.5 M}&\textbf{20 min}\\[1ex]
		\hline
	\end{tabular}}
	\label{tab:ge results}
\end{table}
\begin{table}[h!]
	\caption{Lung opacity score prediction results.}
	\centering
	\resizebox{\columnwidth}{1in}{
    \begin{tabular}{lrrrr}
    \hline
     \noalign{\vskip 2mm}  
           \textbf{Model}     &\multicolumn{1}{c}{\textbf{MAE $\downarrow$}}    & \textbf{PC $\uparrow$} &\textbf{Number of} &\textbf{Training}\\
                   &   &  &\textbf{parameters} &\multicolumn{1}{c}{\textbf{time}}\\[1ex]
        \hline
         \noalign{\vskip 2mm}  
         \textit{COVID-NET}\cite{wong2021towards}&2.249&0.531&12 M&40 min\\
		 \textit{COVID-NET-S}\cite{wong2020covid}&2.227&0.525&12 M&40 min\\
         \textit{ResNet50}\cite{he2016deep} & 1.082& 0.427& 23 M&1.5 hr\\
       \textit{Swin Transformer}\cite{liu2021swin}&0.811 & 0.692& 29 M&2 hr   \\
        \textit{XceptionNet}\cite{chollet2017xception}&0.771 &0.696& 23 M&1.5 hr  \\		
        \textit{InceptionNet}\cite{szegedy2015going}  & 0.614 &0.825& 24 M&1.5 hr \\
       \textit{Feature Extraction}\cite{cohen2020predicting}& 0.881 & 0.701& 20 M&1 hr \\
       \textit{MobileNetV3}\cite{mobilenetv3} &0.741&0.731&4.2 M& 40 min\\
        \textit{\textbf{ViTReg-IP (ours) } }& \textbf{0.512}&\textbf{0.855}&\textbf{5.5 M}&\textbf{20 min}\\[1ex]
        \hline
	\end{tabular}}
	\label{tab:lo results}
\end{table}

\begin{center}
	\begin{table*}[ht!]
		\caption{Results of ViTReg-IP model intra-evaluation.}
    	\centering
		\begin{tabular}{llccccc}
            \hline
            \noalign{\vskip 0.5mm} 
            \multicolumn{1}{l}{\textbf{Data}}&\multicolumn{1}{l}{\textbf{Score}}&\multicolumn{1}{c}{\textbf{Original}}&\multicolumn{1}{c}{\textbf{Training}}&\multicolumn{1}{l}{\textbf{Test}}&\multicolumn{1}{c}{\textbf{MAE $\downarrow$}}& \multicolumn{1}{c}{\textbf{PC $\uparrow$}}\\
            &&\multicolumn{1}{c}{\textbf{Training Size}}&\multicolumn{1}{c}{\textbf{Size*}}&\multicolumn{1}{c}{\textbf{Size}}&&\\
            \hline   
            \noalign{\vskip 0.5mm}
            \textit{Brixia}&Brixia Score &4695&4695&250 & 0.981& 0.622\\
            \textit{Brixia}&Brixia Score &4695&9390&250 & 0.811& 0.763\\
            \textit{RALO}&LO&1878&1878&495 &0.881&0.681\\
            \textit{RALO}&LO&1878&5634&495&0.512&0.855\\
            \textit{RALO}&GE&1878&1878&495&0.931&0.803\\
            \textit{RALO}&GE&1878&5634&495&0.596&0.923\\
            \textit{Danilov et al. COVID-19}&COVID Score&1225&1225&139& 0.389 & 0.951\\[0.5ex]
			\hline
            \multicolumn{7}{l}{\scriptsize{*if combined lung and score replacement is applied}}
		\end{tabular}
	\label{tab:intra results}
	\end{table*}
\end{center}
\begin{center}
	\begin{table*}[ht!]
		\caption{Results of ViTReg-IP model cross-evaluation.}
    	\centering
		\begin{tabular}{lllccccc}
            \hline
            \noalign{\vskip 0.5mm}
            \multicolumn{1}{l}{\textbf{Training}}&\multicolumn{1}{l}{ \textbf{Test}}&\multicolumn{1}{l}{\textbf{Score}}&\multicolumn{1}{c}{\textbf{Original}}&\multicolumn{1}{c}{\textbf{Training}}&\multicolumn{1}{l}{\textbf{Test}}&\multicolumn{1}{c}{\textbf{MAE}} & \multicolumn{1}{c}{\textbf{PC $\uparrow$}}\\
            \multicolumn{1}{l}{\textbf{Data}}&\multicolumn{1}{l}{ \textbf{Data}}&&\multicolumn{1}{c}{\textbf{Training Size}}&\multicolumn{1}{c}{\textbf{Size*}}&\multicolumn{1}{c}{\textbf{Size}}&& \\
            \hline   
            \noalign{\vskip 0.5mm}
            \textit{Brixia}&\textit{Cohen COVID-19} &Brixia Score& 4695&4695&192& 1.86 & 0.461\\
            \textit{Brixia}&\textit{Cohen COVID-19} &Brixia Score& 4695&9390&192& 1.23 & 0.587\\
         	\textit{RALO} &\textit{Cohen COVID-19} & LO&1878&5634& 94 &0.857 &  0.697\\		
         	\textit{RALO} &\textit{Cohen COVID-19} & GE&1878&5634& 94& 0.838 &0.842\\
         	\hline
            \multicolumn{8}{l}{\scriptsize{*if combined lung and score replacement is applied}}\\[0.5ex]
		\end{tabular}
	\label{tab:cross results}
	\end{table*}
\end{center}
To obtain a model with high generalizability, we trained our ViTReg-IP with different combinations of datasets. Depending on the type of data the model was trained with, the results may look different. The experiments included both intra- and cross-evaluation methods.  For the intra-evaluation, in addition to the
RALO dataset, the datasets of Brixia and Danilov \textit{et al.} COVID-19 were used. In each case, the images from the same dataset are split into training and test data and the results of the performance of our trained ViTReg-IP model are collected.
The  data splitting and the results  of  intra-evaluation  are shown in Table \ref{tab:intra results}To avoid biased performance and to ensure the generalizability of the model, cross-evaluation is tested. Splitting the data into training data from one dataset and test data from another dataset is called cross-evaluation. To avoid any biased performance and confirm the generalizability of the model, cross-evaluation is tested. Several tests were performed by training our ViTReg-IP model with different combinations of datasets. The cross-evaluation results are shown in Table \ref{tab:cross results}. In both intra- and cross-evaluation, experiments are performed on images with and without combined lung and score replacement. This augmentation method can only be applied to data that have separate scores for individual lungs, as in the case of the RALO and Brixia datasets. From  Table \ref{tab:cross results} we can see that the performance of the cross-evaluation was lower than that of the intra-evaluation.
\subsubsection{Qualitative Analysis}
\begin{figure}[h!]
 \includegraphics[width=\columnwidth]{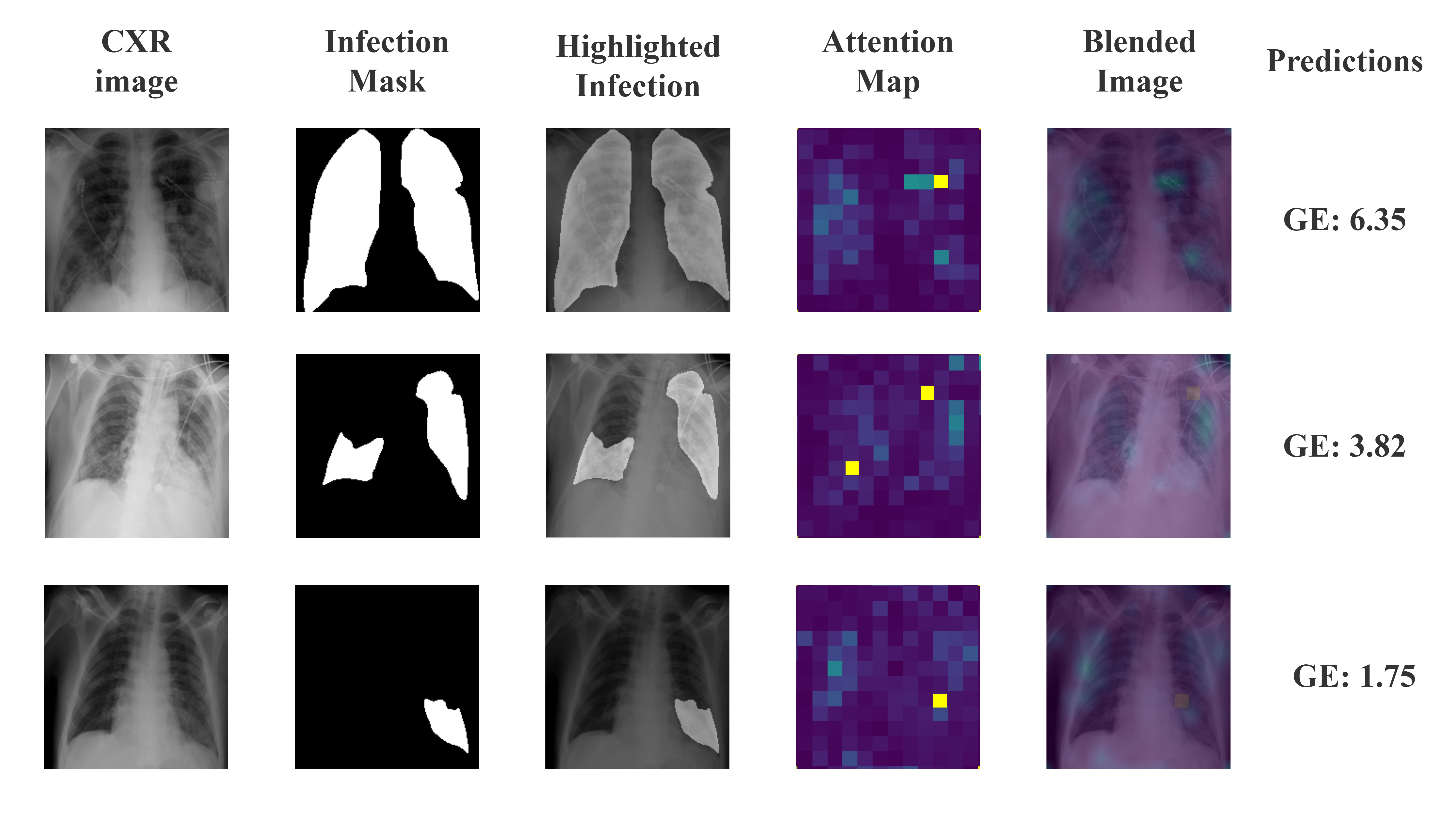}
\centering
\caption{ Attention map of CXRs produced by our ViTReg-IP model for GE score.}
\label{fig: att maps}
\end{figure}

\begin{figure}[h!]
\begin{center}
 \subfloat[CXRs in ascending order of severity scores.]{\centering\includegraphics[width=2in]{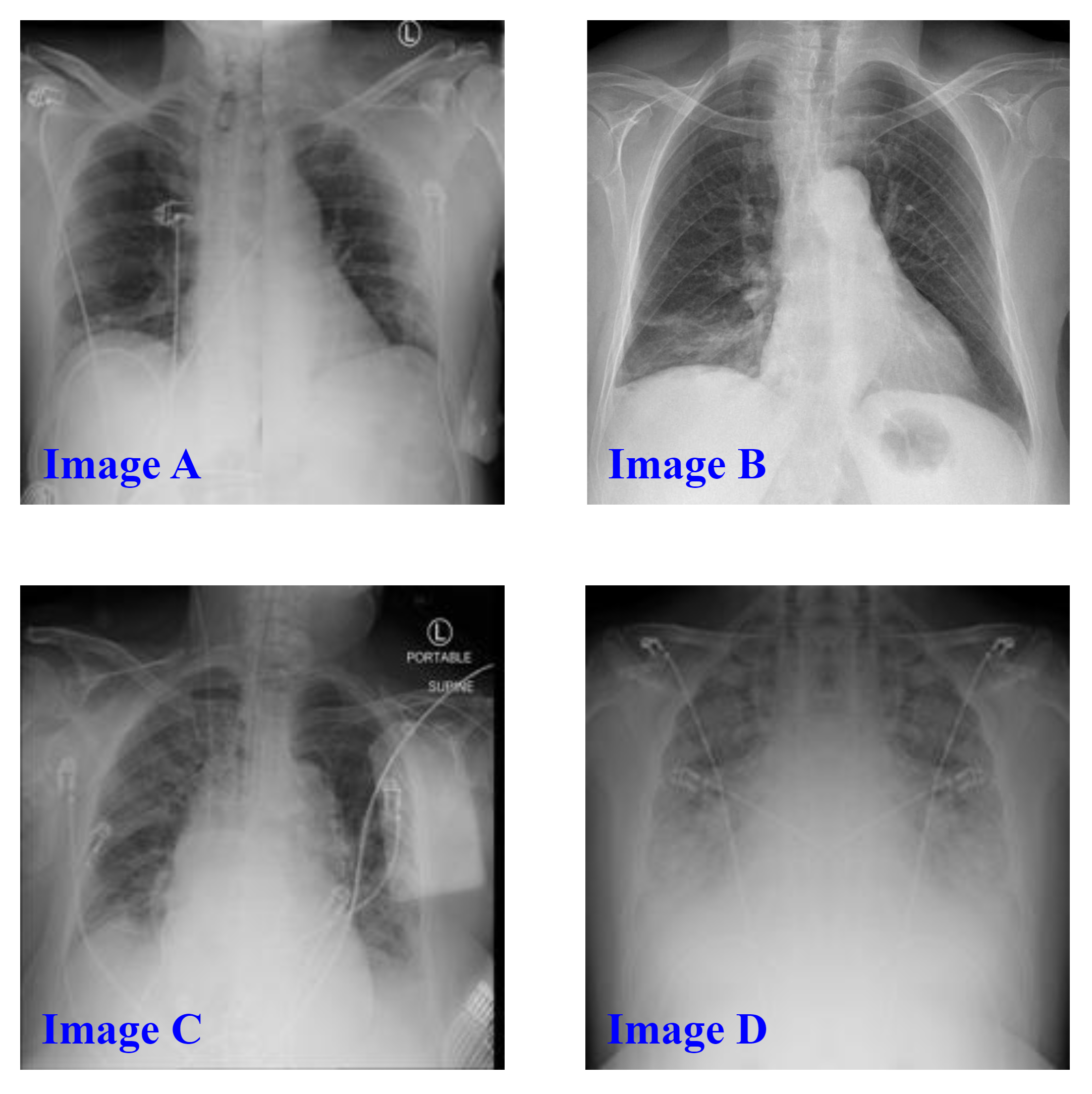} \label{fig: CXR pred}}\\
 \subfloat[The predictions of the severity scores by nine different models. The ground truth scores are shown in the first row of the table.]{\resizebox{\columnwidth}{0.9in}{\begin{tabular}{lcc|cc|cc|cc}
        &\multicolumn{8}{c}{\textbf{Prediction}} \\
     &\multicolumn{2}{c}{\textbf{Image A}}&\multicolumn{2}{c}{\textbf{Image B}}&\multicolumn{2}{c}{\textbf{Image C}}&\multicolumn{2}{c}{\textbf{Image D}}\\
    \hline
     \noalign{\vskip 2mm}  
                    \textbf{Model} & \textbf{GE} & \textbf{LO}
                   & \textbf{GE} & \textbf{LO}
                   & \textbf{GE} & \textbf{LO}
                   & \textbf{GE} & \textbf{LO}\\[1ex]
        \hline
         \textcolor{blue}{\textbf{Ground Truth}}&\textcolor{blue}{\textbf{0}}&\textcolor{blue}{\textbf{0}}&\textcolor{blue}{\textbf{1}}&\textcolor{blue}{\textbf{1.5}}&\textcolor{blue}{\textbf{6}}&\textcolor{blue}{\textbf{3}}&\textcolor{blue}{\textbf{7.5}}&\textcolor{blue}{\textbf{8}}\\
         \hline
         \textit{COVID-NET} \cite{wong2021towards}&1.78&1.54&2.01&2.95&2.13&1.52&5.51&4.86\\
		 \textit{COVID-NET-S} \cite{wong2020covid}&2.01&1.82&2.16&3.12&3.21&2.15&5.13&5.65\\
         \textit{ResNet50} \cite{he2016deep}&0.97& 1.21&1.95&3.14&4.80&2.06&6.92&7.01\\
        \textit{Swin Transformer} \cite{liu2021swin}& 1.06&0.89&0.567&0.75&4.91&2.54&7.91&6.56\\
        \textit{XceptionNet} \cite{chollet2017xception}&0.67&1.12&1.57&1.62&5.58&2.52&7.12&6.78\\		
        \textit{InceptionNet} \cite{szegedy2015going}&0.98&0.99 &0.53&1.23&6.56&3.51&6.88&6.96\\
       \textit{Feature Extraction} \cite{cohen2020predicting}&1.12 &1.01 &1.58&0.94&6.84&3.78&7.95&7.15\\
         \textit{MobileNetV3} \cite{mobilenetv3}&0.91&0.33&1.61&1.21&5.06&3.29&7.44&7.16\\
        \textit{\textbf{ViTReg-IP (ours) } } &\textbf{0.36}&\textbf{0.27}&\textbf{1.05}&\textbf{1.47}&\textbf{5.48}&\textbf{3.12}&\textbf{7.53}&\textbf{7.96}\\[0.5ex]
        \hline
        \noalign{\vskip 2mm}  
	\end{tabular}}
    \label{tab:pred results}}
\centering
\caption{Examples of predictions by the tested models are shown. The CXR images are shown in \ref{fig: CXR pred}. The ground truth and
predicted scores for both Geographic Extent and Lung Opacity are included in Table \ref{tab:pred results}.}
\label{fig: preds}
\end{center}
\end{figure}

\begin{figure}[htbp]
 \subfloat[Learning Curves for training using Geographic Extent (GE).]
 {\includegraphics[width=\columnwidth]{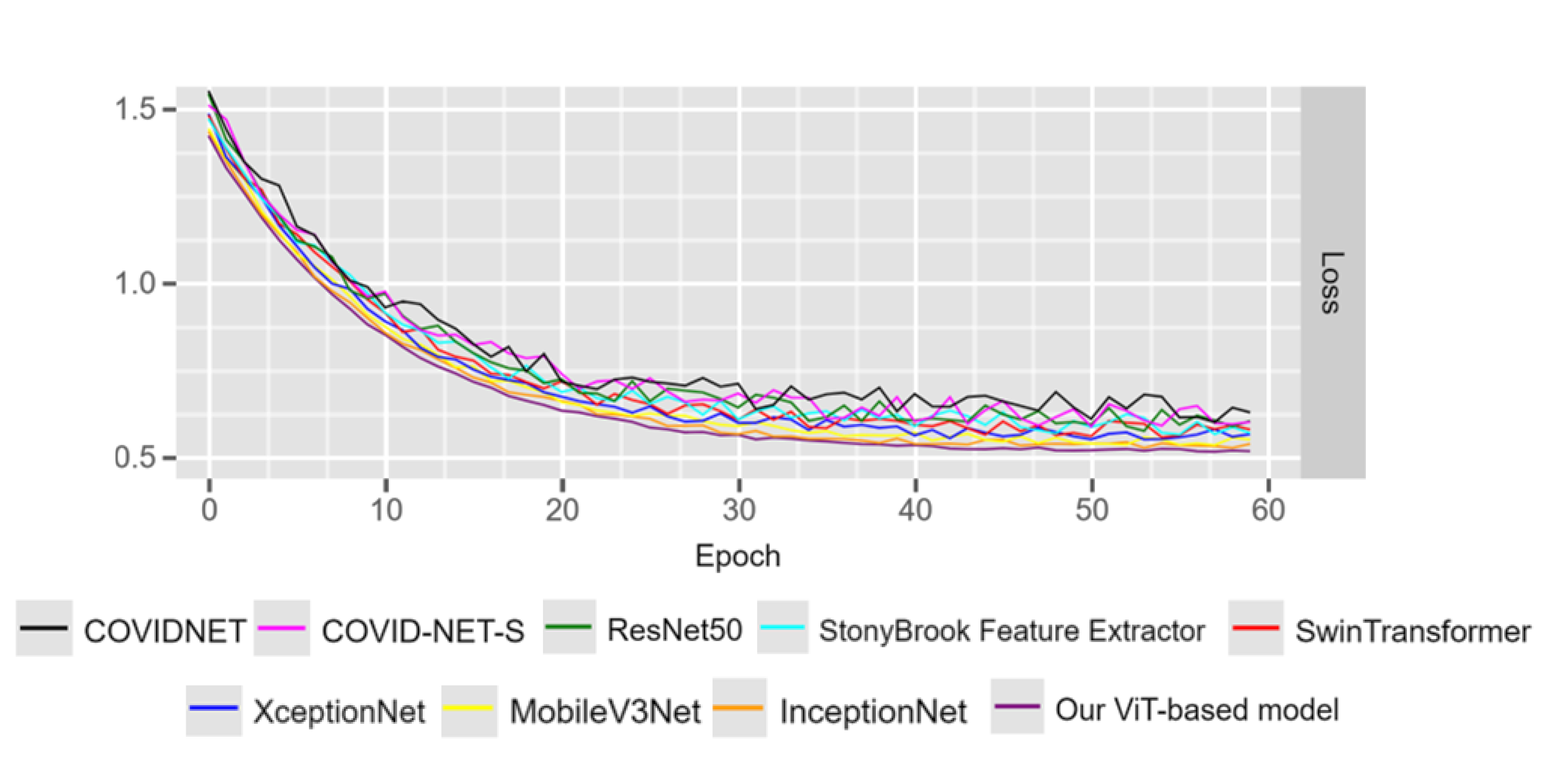}\label{fig: LR ge}}
 \quad
  \subfloat[Learning Curves for training using Lung Opacity (LO).]
  {\includegraphics[width=\columnwidth]{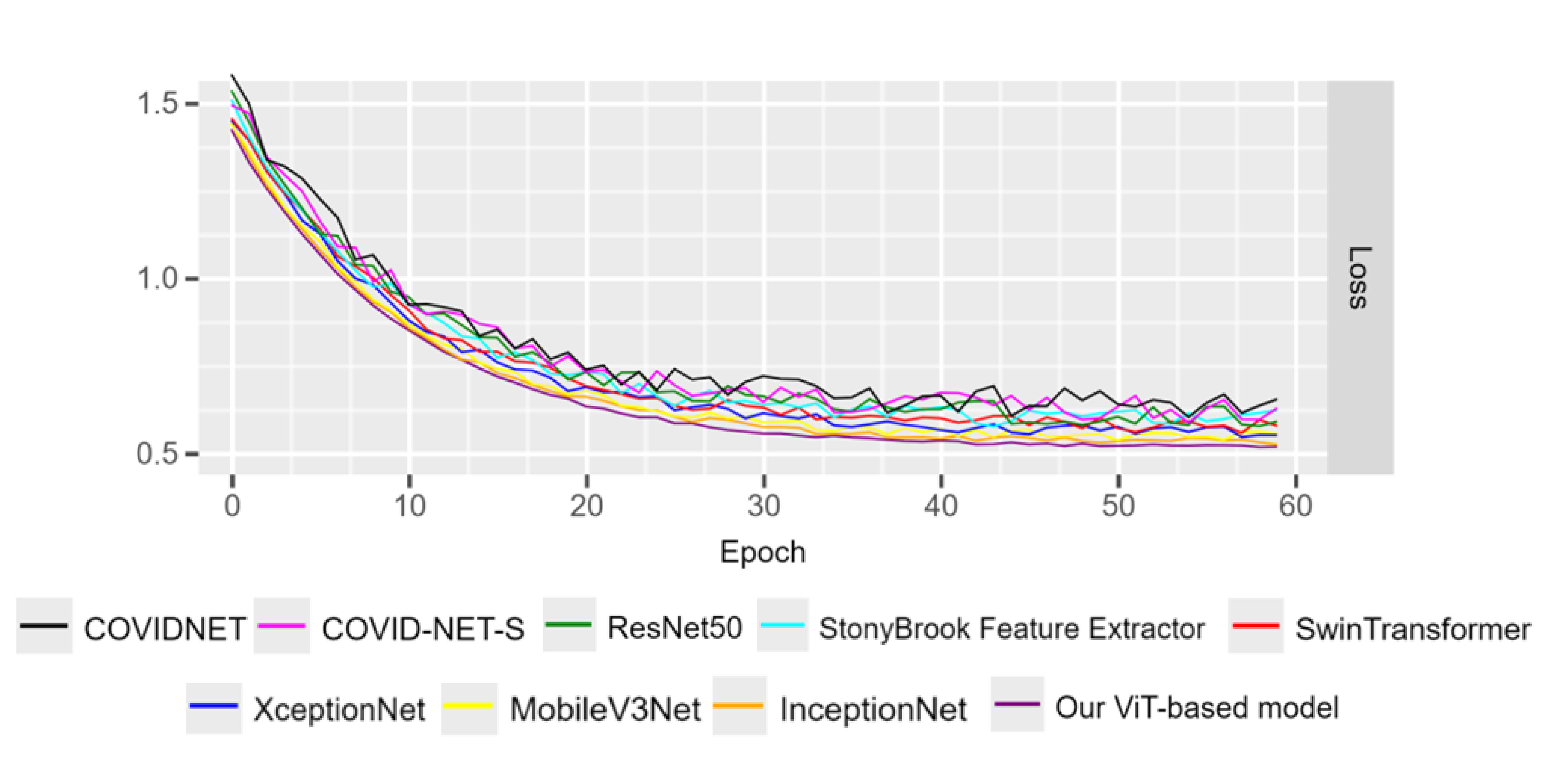}\label{fig: LR lo}}
\centering
\caption{ Learning Curves for the training of the models.}
\label{fig: LR}
\end{figure}

We projected the attention maps to demonstrate the effectiveness of our model in identifying areas at risk of infection. Figure \ref{fig: att maps} shows the ground truth as well as the high-intensity areas corresponding to infection, represented as a feature map. Without the use of sophisticated methods, our recommended model provided a good representation of infection when the ViTReg-IP is trained using geographic extent as a label, with the score correlated to the location of infection. Figure \ref{fig: att maps} shows some examples of the data collected to evaluate the effectiveness of the proposed method for representing lung infection areas. Since the datasets used in this work do not have a ground truth mask for infections, we used CXRs from the QaTa COV19 dataset \cite{QaTa}. In Figure \ref{fig: att maps}, the first column shows the original CXR, the second column shows the actual ground truth of the infection area, the third column shows the image overlaid with the ground truth, column four shows the corresponding attention map, and column five shows a preview of the overlay of the original image with the attention map. The predicted geographic extent values are also included in the last column. The obtained attention maps and predicted scores are highly correlated with the actual location of infection.  In addition, the predicted GE scores also correlate with the extent of infection. This indicates that our proposed model has high efficiency in localizing the area of infection with respect to the high intensities in the spatial attention map.
On the other hand, Figure \ref{fig: preds} shows the predictions made for four CXR images using the different deep learning architectures. The images were selected to have different ground truth scores from the total range to prove that our proposed model is efficient in the whole range of scores. As shown in the table embedded in Figure \ref{fig: preds}, the scores predicted by our proposed model are closest to the ground truth labels of the CXR images as annotated by radiology experts. Even in the absence of infection (scores = 0), as in Image A, the predictions for geographic extent and lung opacity are close to zero compared with the other models. Similarly, the error between the predicted and actual values is smallest when using the ViTReg-IP model in images B, C, and D.
Moreover, Figure \ref{fig: LR} shows the training performance of the proposed model and the eight \textit{state-of-the-art} models presented over 60 epochs. It can be seen that all the trained models converge. As the number of epochs increases, the training loss with the proposed model reaches its stable value in the shortest time compared to the other models. The learning curves are shown for both the geographic extent (Figure \ref{fig: LR ge})  and the lung opacity (Figure \ref{fig: LR lo}) scores.

\subsubsection{Quantitative Analysis}
\begin{figure}[hbt!]
\centering
\subfloat[Model prediction vs ground truth for geographic extent. ]
{\includegraphics[width=1.4in]{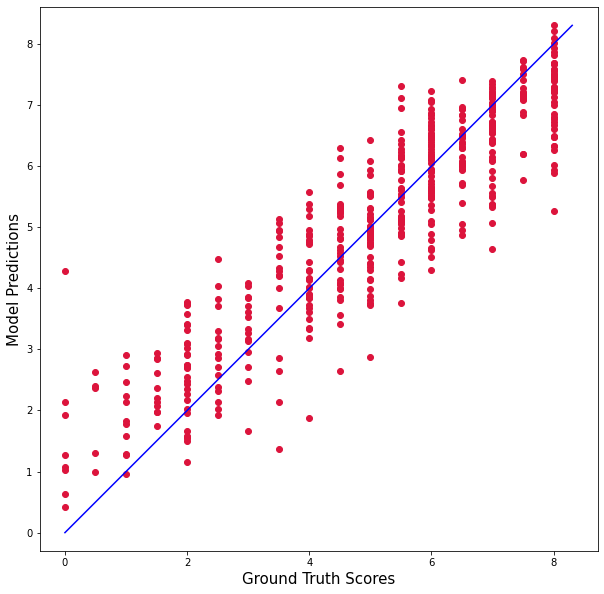}\label{fig: ge pred vs gt}}
\quad
\subfloat[Model prediction vs ground truth for lung opacity.]
{\includegraphics[width=1.4in]{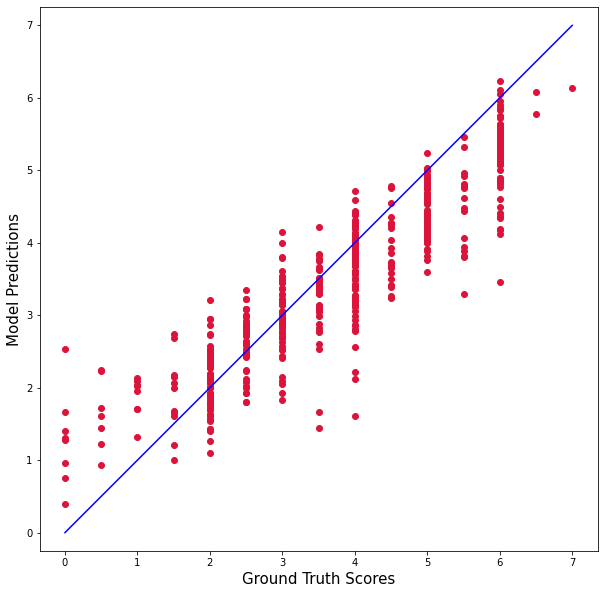}\label{fig: lo pred vs gt}}\\
\centering
\subfloat[Histogram of the absolute errors for geographic extent.]{\includegraphics[width=1.65in]{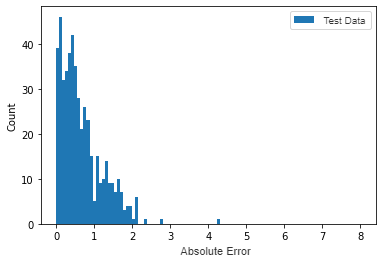}\label{fig: hist geo}}
\quad
\subfloat[Histogram of the absolute errors for lung opacity.]{\includegraphics[width=1.65in]{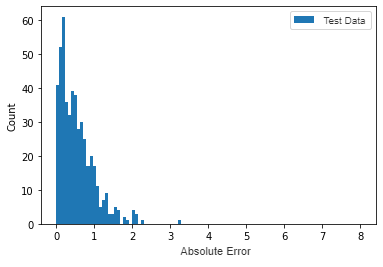}\label{fig: hist lo}}\\
\centering
\caption{VitReg-IP evaluations performed on the test subset.}
    \label{fig: pred vs gt}
    \end{figure}

\begin{figure}[hbt!]
\begin{center}
\subfloat[CMC of the test results for geographic extent.]{\includegraphics[width=1.65in]{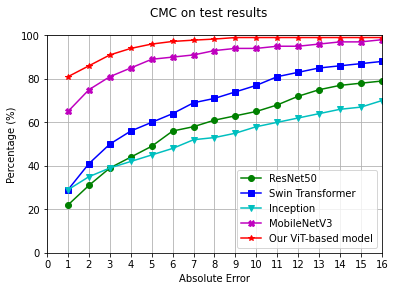}\label{fig: cmc ge}}
\quad
\subfloat[CMC of the test results for lung opacity.]{\includegraphics[width=1.65in]{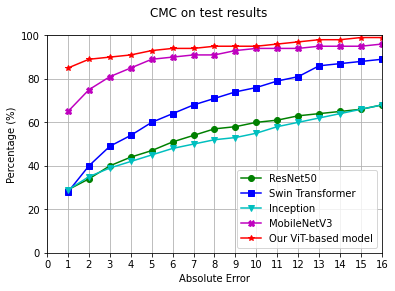}\label{fig: cmc lo}}\\
\end{center}
\caption{The CMC of few tested models.}
\label{fig: cmc}
\end{figure}

The quantitative outcomes of the proposed model are shown in Tables   \ref{tab:ge results} and  \ref{tab:lo results}. In terms of ground truth versus prediction, our model has obtained the best results. The predicted values are relatively close to the actual values, as can be seen in Figure \ref{fig: pred vs gt} for the RALO dataset. The same is true for both annotations, i.e., geographic extent and lung opacity scores. We also plotted the histograms of the absolute errors obtained with the  test images (Figures \ref{fig: hist geo} and \ref{fig: hist lo}).  From these histograms, we can see that the highest bars are shifted to the left meaning that a large number of test images have a small prediction error. It can be seen that most of the errors of the individual test images are in the range of 0-1, giving the total error. 
We also considered the cumulative matching curves (CMC) of some tested models to evaluate their performance. The curves for test images for both scores are shown in Figure \ref{fig: cmc}. Our proposed model outperformed three other models. Each color represents one model. For the GE score (Figure \ref{fig: cmc ge}), about 80\% of the test images in our model have a prediction error below the first error threshold (here it is set to one). The other models tested, such as RestNet50, Swin Transformer, Inception, and MobileNetV3, resulted in a much lower percentage for the same threshold. Similar behavior was obtained with the CMC of the LO score (Figure \ref{fig: cmc lo}).   
\subsubsection{Ablation Studies}
We performed a series of ablation studies to better understand the contributions of each component in our ViTReg-IP model. As shown in Tables \ref{tab:loss results}, \ref{tab:opt results}, \ref{tab:fc results},  \ref{tab:aug results},   \ref{tab:on-aug results},  and \ref{tab:seg}.
First, we conducted an ablation study to determine the impact of the loss function used on the performance of our proposed model. The model is trained using the CXR images for each loss function and the results are previewed. The loss functions used are the Huber loss, the MSE loss, and the smooth L1 loss in addition to the L1 loss. The results are shown in Table \ref{tab:loss results}.  The Huber loss function uses a quadratic term to create a criterion if the absolute  error is less than a given parameter, otherwise, a scaled L1 term is used. The MSE loss establishes a standard that evaluates the mean squared error between the predicted value and the target value. The smooth L1 loss uses a quadratic  term if the absolute  error is less than a given parameter, and an L1 term otherwise.
The results show that using L1 Loss as the loss function gives the best results. Using L1 Loss, MAE is smallest with values of 0.569 and 0.512 and PC with highest values of 0.923 and 0.855 for geographic extent and lung opacity, respectively. 
\begin{center}
	\begin{table}[h!]
		\caption{Ablation study results for loss function performance.}
    	\centering
		\resizebox{2.5in}{0.4in}{
        \begin{tabular}{lcccc}
            \hline
            \noalign{\vskip 0.5mm}  
              & \multicolumn{2}{c}{\textbf{GE}}& \multicolumn{2}{c}{\textbf{LO}}\\
              \textbf{Loss Function}& \textbf{MAE$\downarrow$}&\textbf{PC $\uparrow$}&\textbf{MAE$\downarrow$}&\textbf{PC $\uparrow$}\\
            \hline   
            \noalign{\vskip 0.5mm}  
          	 L1Loss &\textbf{0.569}&\textbf{0.923}&\textbf{0.512}&\textbf{0.855}\\		
             MSE Loss&0.590&0.917&0.612&0.817\\
             Smooth L1 Loss&0.615&0.913&0.542&0.843\\
             Huber Loss&0.637&0.909&0.601&0.807\\ [0.5ex]
			\hline
		\end{tabular}}
	\label{tab:loss results}
	\end{table}
\end{center}

To choose the best optimizer for our model, we trained our ViTReg-IP with five different optimizers and compared the results in terms of MAE and PC. The optimizers tested include Adadelta, SGD, Adam, AdamW, and RMSprop. Table \ref{tab:opt results} shows the results of the tests and shows that SGD ensures the best performance. 
\begin{center}
	\begin{table}[h!]
		\caption{Ablation study results for optimizer performance.}
    	\centering
		\resizebox{2.5in}{0.5in}{
        \begin{tabular}{lcccc}
        	\hline
            \noalign{\vskip 0.5mm}  
              & \multicolumn{2}{c}{\textbf{GE}}& \multicolumn{2}{c}{\textbf{LO}}\\
              \textbf{Optimizer}& \textbf{MAE $\downarrow$}&\textbf{PC $\uparrow$}&\textbf{MAE $\downarrow$}&\textbf{PC $\uparrow$}\\
            \hline   
            \noalign{\vskip 2mm}  
          	 SGD &\textbf{0.569}&\textbf{0.923}&\textbf{0.512}&\textbf{0.855}\\		
             Adadelta&0.743&0.881&0.667&0.811\\
             Adam&0.885&0.841&0.939&0.613\\
             AdamW&0.901&0.821&0.813&0.691\\
            RMSprop&1.178&0.697&0.909&0.618\\[0.5ex]
			\hline
		\end{tabular}}
	\label{tab:opt results}
	\end{table}
\end{center}

Table \ref{tab:fc results} shows the ablation study performed for the size of the linear fully connected layer connected to the output of the transformer in the regressor of our ViTReg-IP. We tested a range of sizes and previewed the MAE and PC values corresponding to each test. The results show that the 128 FC layer we chose gives the best results compared to other sizes.
\begin{center}
	\begin{table}[h!]
		\caption{Ablation study results for FC layer size performance.}
    	\centering
		\resizebox{2.5in}{0.6in}{
        \begin{tabular}{rcccc}
            \hline
            \noalign{\vskip 0.5mm}  
              & \multicolumn{2}{c}{\textbf{GE}}& \multicolumn{2}{c}{\textbf{LO}}\\
             \multicolumn{1}{l}{\textbf{FC Size}}& \textbf{MAE $\downarrow$}&\textbf{PC $\uparrow$}&\textbf{MAE $\downarrow$}&\textbf{PC $\uparrow$}\\
            \hline   
            \noalign{\vskip 0.5mm}  
          	 50&0.663&0.922&0.563&0.845\\
             75&0.662&0.921&0.584&0.839\\
             100&0.686&0.910&0.556&0.845\\
             \textbf{128}&\textbf{0.569}&\textbf{0.923}&\textbf{0.512}&\textbf{0.855}\\
             150&0.649&0.901&0.546&0844\\
             175&0.646&0.902&0.529&0.849\\ [0.5ex]
			\hline
		\end{tabular}}
	\label{tab:fc results}
	\end{table}
\end{center}

The next study targeted the effect of the augmentation methods on model performance. Training of our ViTReg-IP was performed using either combined lung and score replacement or score-correlated CutMix, both, or neither methods. The results in Table \ref{tab:aug results} show that combined lung and score replacement made a greater contribution to improving model performance, with MAE decreasing the most and PC increasing the most when applied alone, compared to score-correlated CutMix applied alone. Score-correlated CutMix also improved results, but to a lesser extent.
\begin{center}
	\begin{table*}[h!]
		\caption{Ablation study results for augmentation performance.}
    	\centering
        \begin{tabular}{cccccc}
            \hline
            \noalign{\vskip 0.5mm}  
             \multicolumn{2}{c}{\textbf{Augmentation}}& \multicolumn{2}{c}{\textbf{GE}}& \multicolumn{2}{c}{\textbf{LO}}\\
             \multicolumn{1}{c}{\textbf{Combined Lung \& Score Replacement}}&\multicolumn{1}{c}{\textbf{Score-correlated CutMix}}&\textbf{MAE $\downarrow$}&\textbf{PC $\uparrow$}&\textbf{MAE $\downarrow$}&\textbf{PC $\uparrow$}\\
             \hline   
            \noalign{\vskip 0.5mm}  
          	 $\times$ & $\times$  &1.032&0.778&0.926&0.635\\
           $\checkmark$ & $\times$&0.655&0.905&0.573&0.843\\		
            $\times$& $\checkmark$&0.931&0.803&0.881&0.681\\
          $\checkmark$&$\checkmark$ &\textbf{0.569}&\textbf{0.923}&\textbf{0.512}&\textbf{0.855}\\[0.5ex]
			\hline
		\end{tabular}
	\label{tab:aug results}
	\end{table*}
\end{center}

The study revealed in Table \ref{tab:on-aug results} considers several online augmentations of the \textit{state-of-the-art}. It was conducted to confirm that choosing CutMix as the online augmentation step produced the best results. CutOut replaces a random box from each image with a black one \cite{cutout}, Attentive CutMix replaces the most  descriptive regions of an image based on the intermediate attention maps of a feature extractor with those of another image \cite{attcutmix}, and MixUp performs a fusion of two images to create a new image \cite{mixup}.  GridMix uses patch-level label prediction for local context mapping and grid-based mixing \cite{gridmix}. SuperPixelMix uses information merging to create a new style of image augmentation based on superpixel decomposition \cite{hammoudi2023superpixelgridmasks}. PuzzleMix is a MixUp approach that directly uses saliency data and supporting statistics \cite{puzzlemix}. TransMix is similar to CutMix in terms of mixing images, however, it blends labels based on the Vision Transformers attention matrices \cite{transmix}. Horizontal image flipping and image blurring are two traditional augmentation methods that were also tested \cite{buslaev2020albumentations}. For all tested augmentation methods that use image mixing, the scoring strategy explained in \eqref{eq5} is applied for each case.
All previously mentioned augmentation methods are tested. As confirmed by the test results in terms of the lowest MAE and the highest PC, the best results were obtained using CutMix as the online data augmentation.

To investigate whether lung segmentation \cite{Mansoor2020} can improve our model or not, we tested some segmentation architectures  to segment the lung regions from the original CXR image as a preliminary step before training the regression model.  Thus, lung segmentation can be considered a preprocessing step for the CXR input images. With respect to \textit{state-of-the-art} segmentation architectures, we used MA-Net \cite{MA-net}, PAN \cite{ PAN }, and UNet \cite{unet}. We decided to test a traditional CNN-based model in addition to our ViTReg-IP. Both models were trained with both geographic extent and lung opacity scores. The results in Table \ref{tab:seg} show that similar results are obtained whether or not segmentation is performed before training our proposed model. In addition, we tested ResNet50 as a backbone for the regressor with or without lung segmentation for the CXR input data. The results in terms of MAE and PC are shown in Table \ref{tab:seg}. As can be seen, the MA-Net lung segmentation method improved the performance of the ResNet50-based model.
\begin{center}
	\begin{table}[h!]
		\caption{Ablation study results for online augmentation performance.}
    	\centering
		\resizebox{\columnwidth}{1in}{\begin{tabular}{lcccc}
            \hline
            \noalign{\vskip 0.5mm}  
             \multicolumn{1}{c}{\textbf{Online}}& \multicolumn{2}{c}{\textbf{GE}}& \multicolumn{2}{c}{\textbf{LO}}\\
             \multicolumn{1}{c}{\textbf{Augmentation}} &\textbf{MAE $\downarrow$}&\textbf{PC $\uparrow$}&\textbf{MAE $\downarrow$}&\textbf{PC $\uparrow$}\\
            \hline   
            \noalign{\vskip 0.5mm}
            \textbf{Score-correlated CutMix}&\textbf{0.569}&\textbf{0.923}&\textbf{0.512}&\textbf{0.855}\\
            TransMix\cite{transmix}&0.582&0.921&0.551&\textbf{0.855}\\
            SuperPixelMix\cite{hammoudi2023superpixelgridmasks}&0.789&0.892&0.712&0.873\\
            Horizontal Flip \cite{buslaev2020albumentations}&0.599&0.915&0.574&0.844\\
            Blur \cite{buslaev2020albumentations}&0.602&0.919&0.547&0.843\\
            MixUp\cite{mixup}&0.611&0.904&0.651&0.837\\
            CutOut\cite{cutout} &0.642&0.914&0.601&0.814\\		
            Attentive CutMix\cite{attcutmix}&0.832&0.889&0.789&0.795\\
            PuzzleMix\cite{puzzlemix}&0.754&0.851&0.721&0.732\\            
            GridMix\cite{gridmix}&0.848&0.832&0.834&0.701\\[0.5ex]
			\hline
		\end{tabular}}
	\label{tab:on-aug results}
	\end{table}
\end{center}
\begin{center}
	\begin{table*}[ht!]
		\caption{Ablation study results for lung segmentation performance.}
    	\centering
        \begin{tabular}{lcccccccc}
            \hline
            \noalign{\vskip 0.5mm}  
             \multicolumn{1}{c}{}& \multicolumn{4}{c}{\textbf{ViT-Reg-IP (ours)}}& \multicolumn{4}{c}{\textbf{ResNet50}}\\
             \multicolumn{1}{c}{\textbf{Segmentation}}& \multicolumn{2}{c}{\textbf{GE}}& \multicolumn{2}{c}{\textbf{LO}}& \multicolumn{2}{c}{\textbf{GE}}& \multicolumn{2}{c}{\textbf{LO}}\\
              &\textbf{MAE $\downarrow$}&\textbf{PC $\uparrow$}&\textbf{MAE $\downarrow$}&\textbf{PC $\uparrow$}&\textbf{MAE $\downarrow$}&\textbf{PC $\uparrow$}&\textbf{MAE $\downarrow$}&\textbf{PC $\uparrow$}\\
            \hline   
            \noalign{\vskip 0.5mm}
            No segmentation&\textbf{0.569}&\textbf{0.923}&\textbf{0.512}&\textbf{0.855}&1.107&0.684&1.082&0.427\\
            Segmentation: MA-Net \cite{MA-net}&0.578&0.919&0.534&0.841&\textbf{0.798}&\textbf{0.776}&\textbf{0.764}&\textbf{0.762}\\		
            Segmentation: PAN \cite{PAN}&0.589&0.914&0.554&0.831&0.812&0.754&0.809&0.759\\
            Segmentation: UNet \cite{unet}&0.654&0.849&0.612&0.798&0.952&0.521&0.935&0.507\\ [0.5ex]
			\hline
		\end{tabular}
	\label{tab:seg}
	\end{table*}
\end{center}

\section{Analysis of results and  discussions}
\label{Discussion}
In this work, a novel deep-learning method is developed and validated to predict the severity of COVID-19-infected patients. According to the experimental results, the proposed model holds the potential for clinical diagnosis and early therapy by accurately and rapidly predicting the severity of infection in COVID-19 patients using CXR images.
Our proposed ViTReg-IP model can predict accurate quantification of the severity of lung infection, although it was trained on a small dataset. For the evaluation of a system based on quantitative assessments, we proposed an image analysis method. In all experiments, the use of a combined ViT regressor yielded the best performance in score prediction. Compared with the radiologist's clinical annotations for geographic extent or lung opacity on the entire test set of 495 CXRs, the reported mean absolute errors are less than 0.5, with a range of ground truth scores of $[0, 8]$. For a diagnosis of urgency that provides a very accurate assessment of the degree of infection, a MAE of less than 0.5 is considered an acceptable error for the network and radiologists. Other features of the strategy proposed in this study make it superior to similar methods in the literature. The same experiments are performed with other competing architectures.
COVID-NET, COVID-NET-S, and Stonybrook Feature Extraction are trained in addition to training ResNet50, InceptionNet, XceptionNet, Swin Transformer, and MobileNetV3 as a backbone to the regressor instead of the ViT. The results of training the deep learning architectures are compared with the results of our proposed model to demonstrate the value of the work. The proposed model outperforms the existing deep learning models in terms of MAE and PC compared to other supervised AI-based prediction models, as shown in Tables \ref{tab:ge results} and \ref{tab:lo results}.
The experimental results show that the learned ViTReg-IP model, when trained on the analyzed RALO dataset of CXR data, could achieve MAE between the predicted values and the radiologist's scores for the geographic extent of 0.596 and the opacity extent of 0.512, which is the lowest error compared to the \textit{state-of-the-art}. Similarly, the best performance can also be indicated by the PC metric, where it is highest using our proposed model with 0.923 and 0.855 for geographic extent and lung opacity, respectively.

When choosing configuration values, considering training costs is as important as focusing on the absolute best performance. For this reason, it is important to examine the number of parameters and the training time for each model. This insight is taken into account when selecting a model over a time-consuming training process. Tables \ref{tab:ge results} and \ref{tab:lo results} illustrate the computational efficiency indicated by the number of parameters and training time of the proposed model. The proposed model provides the lowest MAE, although it has only 5.5 million parameters that take at most 20 minutes to train, resulting in a low computational cost. On the other hand, the world and the context in which the model is applied are potentially constantly changing. This change may take the form of an urgent pandemic, a change in interest rates, or a model failure. In addition, the model may be applied to a new context: For example, the model may be trained again to predict a different score for a new virus or a particular lung infection. The time required to train a model is one of several factors to consider when selecting a model. Because of data drift, it is rarely possible to train the model only once. The input data and the target variable can change over time, which can cause the model to perform significantly worse if it's not continuously re-trained. In addition, more data may be collected over time that we want to incorporate in order to obtain the best possible model.

In addition, Tables \ref{tab:intra results} and \ref{tab:cross results} show the generalization and robustness of our model over different combinations of CXR images with different labeling scores. These tests include both intra- and cross-validation tests with different combinations of CXR images. The MAE values of the intra-validations performed range from 0.981 to 0.389 and the PC values range from 0.622 to 0.951. These results confirm the ability of our ViTReg-IP to perform well on different data. The results of the cross-tests also indicate a good performance of our model. The MAE values of the cross-tests have a maximum of 1.86 and a minimum PC of 0.461. The results are evidence of the successful application of our proposed model when trained on CXR images from different datasets, and it can still provide efficient results that can be used for primary diagnosis and rapid intervention in emergencies.

The ablation studies performed have highlighted the different contributions of multiple parameters. First, we focused on the loss function chosen in our studies. Table \ref{tab:loss results} shows that the use of L1Loss gave the best results compared to Huber, SmoothL1, and MSE losses. Similarly, we tested the efficiency of choosing the right optimizer and the first FC layer size. The results shown in Tables \ref{tab:opt results} and \ref{tab:fc results} confirm the correct use of SGD optimizer and 128-long FC layer. In addition, data augmentation made a large contribution to improving the performance of our model, with the offline combined lung and score replacement augmentation making the largest contribution. This can be seen in Table \ref{tab:aug results} by testing the performance of the model with and without offline and online augmentation steps. The decision to use CutMix as the online augmentation step was made after testing various \textit{state-of-the-art} augmentation methods. As can be seen in Table \ref{tab:on-aug results}, CutMix, TransMix, and Superpixelmix each provide the best results.

Although lung segmentation is a crucial first step in radiological lung imaging, in which a computer-assisted procedure separates the boundaries of the lung from the surrounding thoracic tissue on CXR images, Table \ref{tab:seg} proves that lung segmentation did not improve the results in our case. The results show that the model works efficiently with or without the segmentation of the input CXR. Even when the segmentation models with the highest performance are applied in separating the lung from the surrounding in CXR images, the robustness of our ViT-based model is not improved compared to using non-segmented images. This can be due to the particular self-attention mechanisms in the layers of the Transformer. However, applying segmentation to the input images before training a traditional CNN such as ResNet50 resulted in improved performance (MAE is 1.107 for GE, 1.082 for LO without segmentation, and 0.798 for GE 0.764 for LO with segmentation using MA-Net). This can be explained by the fact that transformers interpret the images as patches and solve sequence tasks rather than complete images, while CNN compares image features located at approximately the same positions as the original image. Masking a particular region of the CXR by segmentation can deceive the transformer and lead to a larger error in prediction. As can be seen, in contrast to the \textit{state-of-the-art}, segmentation can be bypassed in our proposed model and still achieve good performance. In this case, we can additionally reduce the high computational cost required to train a segmentation model and apply the segmentation to the input datasets.

In summary, disease detection is the focus of the bulk of Deep Learning-based COVID-19 research on diseases. The remaining relevant literature shows that disease severity assessment typically involves binary categorization as opposed to multi-classification of COVID-19. To the best of the authors' knowledge and after a careful review of the literature, the main advantages and contributions of the methodology proposed in this paper are that the proposed study differs from other disease detection studies in that it not only detects COVID-19 disease but also determines the severity of the disease. This study avoids the significant amount of work required for lesion segmentation, which saves time and improves the ability of clinicians to quickly and efficiently perform COVID-19 disease assessment studies based on deep learning methods.
\section{Conclusion}
\label{conclusion}
In this study, we hypothesized that a generalized transformer-based approach could reliably and rapidly predict the degree of pulmonary infection in patients with COVID-19 by exploiting multi-score datasets of graded CXRs and comparing them with ground truth scores annotated by radiologists. Using computer-aided severity evaluation of CXR images from COVID-19-positive patients, the experimental results demonstrate the effectiveness of the proposed model and its potential to be a helpful tool for clinicians and healthcare workers. Our experimental results showed that the proposed network could be stably trained with a small dataset, made comparisons with the \textit{state-of-the-art} methods, and produced results with the lowest error that were highly correlated with radiological results.  Moreover, this open-access approach provides a biomedical assistance tool that may be useful for automatically alerting on CXRs scored by physicians which may require double-checking due to an identified high score deviation. In addition to being highly efficient, our model has the advantage of being computationally inexpensive due to its short training time and not requiring segmentation as a preprocessing step. As a result, the ViTReg-IP can be retrained on new data to respond to the same or different pulmonary infections by predicting the relative corresponding score.

\section*{Author contributions}
\small
\textit{Conceptualization}: Bouthaina Slika, Fadi Dornaika \& Karim Hammoudi; \textit{Methodology}: Bouthaina Slika, Fadi Dornaika \& Karim Hammoudi;\textit{ Formal analysis and investigation}: Bouthaina Slika, Fadi Dornaika \& Karim Hammoudi; \textit{Writing - Review \& Editing}: all authors.
 \bibliographystyle{plain}
 \bibliography{arxiv}
\end{document}